\documentclass[useAMS,usenatbib]{mn2e}

\usepackage{subfig}
\usepackage{graphicx}
\usepackage[countmax]{subfloat}
\usepackage{rotating}
\usepackage{natbib}
\usepackage{amsmath}
\usepackage{longtable}
\usepackage{hyperref}
\usepackage{aas_macros}

\title[Stellar population synthesis models between 2.5 and $5\,\mu m$]
  {Stellar population synthesis models between 2.5 and 5$\,\mu$m based on the empirical IRTF stellar library}
\author[B. R\"{o}ck et al.]
  {B. R\"{o}ck,$^{1,2}$ A. Vazdekis,$^{1,2}$ R.F. Peletier,$^3$ J.H. Knapen,$^{1,2}$ J. Falc\'{o}n-Barroso,$^{1,2}$ \\
  $^1$Instituto de Astrof\'{i}sica de Canarias, Calle V\'{i}a  L\'{a}ctea, s/n, E-38205 La Laguna, Tenerife, Spain\\
  $^2$Departamento de Astrof\'{i}sica, Universidad de La Laguna, E-38205 La Laguna, Tenerife, Spain\\
  $^3$Kapteyn Astronomical Institute, University of Groningen, Postbus 800, 9700 AV, Groningen, The Netherlands}

\date{Released 2013 Xxxxx XX}

\pagerange{\pageref{firstpage}--\pageref{lastpage}} \pubyear{2013}

\def\LaTeX{L\kern-.36em\raise.3ex\hbox{a}\kern-.15em
    T\kern-.1667em\lower.7ex\hbox{E}\kern-.125emX}

\begin{document}

\label{firstpage}

\maketitle

\begin{abstract}
We present the first single-burst stellar population models in the infrared wavelength range between 2.5 and 5$\,\mu$m which are exclusively based on empirical stellar spectra. Our models take as input 180 spectra from the stellar IRTF library.
Our final single-burst stellar population models are calculated based on two different sets of isochrones and various types of initial mass functions of different slopes, ages larger than 1 Gyr and metallicities between ${\rm[Fe/H]=-0.70}$ and ${\rm [Fe/H]=0.26}$. They are made available online to the scientific community on the MILES webpage. 
We analyse the behaviour of the \textit{Spitzer} [3.6]$-$[4.5] colour calculated from our single stellar population models and find only slight dependencies on both metallicity and age. 
When comparing to the colours of observed early-type galaxies, we find a good agreement for older, more massive galaxies that resemble a single-burst population. Younger, less massive and more metal-poor galaxies show redder colours with respect to our models. This mismatch can be explained by a more extended star formation history of these galaxies which includes a metal-poor or/and young population.
Moreover, the colours derived from our models agree very well with most other models available in this wavelength range.
We confirm that the mass-to-light ratio determined in the \textit{Spitzer} ${\rm [3.6] \, \mu m}$ band changes much less as a function of both age and metallicity than in the optical bands.
\end{abstract}


\section{Introduction}

Most galaxies cannot be resolved photometrically into single stars due to their large distances. In order to study their stellar content, observations of their integrated light can be analysed by means of so-called single-burst stellar population (SSP) synthesis models which reproduce a population of a single age and metallicity \citep[e.g.,][]{Vazdekis99, Bruzual03, Maraston05, Maraston09, Vazdekis10, Conroy12}. These models are constructed by populating stellar evolutionary isochrones with stellar spectra following a particular initial mass function (IMF). The necessary spectra originate from large stellar libraries \citep[e.g.,][]{Lejeune97, Lejeune98, Lancon00, Westera02, Prugniel01, LeBorgne03, Valdes04, Sanchez06, Rayner09}. These are compilations of either observed or theoretically constructed stellar spectra, ideally covering a large range of the stellar atmospheric parameters effective temperature $T_{\text{eff}}$, surface gravity $\log(g)$ and metallicity [Fe/H]. Whereas a number of empirical stellar libraries exist in the optical wavelength range \citep[e.g.,][Pickles, ELODIE, STELIB, Indo-US, MILES]{Pickles98, Prugniel01, LeBorgne03, Valdes04, Sanchez06}, only very few empirical libraries are available in the near-infrared (NIR) and infrared (IR) wavelength range \citep[e.g.,][]{Lancon00}. Throughout this paper, we will refer to the wavelength range until the end of the K-band as NIR, and as IR to the range between 2.5 and 5$\,\mu$m, in accordance with the generally used nomenclature. Most SSP models employ theoretical stellar libraries \citep[e.g.,][BaSeL]{Kurucz92, Lejeune97, Lejeune98, Westera02} in this wavelength range. These are, however, generally hampered by low resolution and uncertain prescriptions of theoretical model atmospheres. On the other hand, the caveats of empirical stellar libraries are their natural limits in parameter coverage. Since most of the stars of the stellar libraries are observed in the Solar neighbourhood, their metallicity coverage is rather small around the Solar value. However, the quality of the resulting SSP models depends very much on the parameter coverage of the input stellar libraries. The NASA Infrared Telescope Facility (IRTF) spectral library \citep{Cushing05, Rayner09} is the only empirical library available at the moment that extends towards the IR with a reasonably good stellar parameter coverage (see Section \ref{IRTF}). 

There are various ways in which observed galaxies can be studied with the help of SSP models. One possibility is to compare their observed colours with models. Another option is to focus on absorption line strength indices visible in both the recorded and the modelled spectra \citep[e.g.,][]{Worthey92, Vazdekis96, Vazdekis03, Thomas03, Thomas05, Cervantes09, Thomas10} or even carrying out a full spectrum fitting approach \citep[see e.g.][]{CidFernandes05, Koleva09}. In the latter case, the whole spectral energy distribution of an observed unresolved stellar population is reproduced by a best-fitting combination of model spectra. This way, valuable constraints on the underlying stellar populations and their star formation histories (SFH) can be drawn.

The general lack of observations of both stars and galaxies in the NIR and IR wavelength range gets also reflected by the small number of models available so far in this regime. Models like the Galaxy Evolutionary (GALEV) synthesis models \citep{Kotulla09}, the Flexible Stellar Population Synthesis (FSPS) models \citep{Conroy09}, the models of \citet{Bruzual03}, the models of \citet{Maraston05} and others which are based on the theoretical BaSeL-library \citep{Lejeune97, Lejeune98, Westera02} are the only ones providing model spectra up to the IR. The latter ones by \citet{Bruzual03} are also available based on the empirical STELIB library \citep{LeBorgne03} for the optical wavelength range. A number of other models exist in the NIR wavelength range up to the end of the $K$ band. For example, \citet{Mouhcine02} calculated models in this regime based on the empirical stellar library of \citet{Lancon00}. Since this library includes only cool stars, they added theoretical hot stars. A similar approach was carried out by \citet{Maraston05} who also took advantage of the empirical spectra of the library by \citet{Lancon00} for cool stars. The set of models derived by \citet{Maraston09} is completely based on the empirical low resolution stellar library by \citet{Pickles98}. Recently, Meneses-Goytia et al. (2014) took advantage of the IRTF-library to create SSP models ranging up to ${\rm2.4 \,\mu}$m. Their approach is quite similar to the one conducted by us, since the input stellar library and also the routines used to create the model spectra are identical. Here, however, we focus on different wavelength ranges and pursue different global aims (details see below). 

The main goal of our work as presented in this paper is to extend the SSP models developed in the optical spectral range between ${\rm \lambda=3465 \, \AA}$ and ${\rm \lambda=9469 \,\AA}$ by \citet{Vazdekis03, Vazdekis10, Vazdekis12} to the IR wavelength regime. The IR spectral range offers certain advantages compared to the optical. Firstly, it is much less affected by dust extinction. Moreover, IR light traces cool, old stars which dominate the baryonic mass in galaxies. Finally, these wavelengths are also very suitable to quantify the contribution of asymptotic giant branch (AGB) stars.\looseness-2

Our models are based on the same modelling routines (see Section \ref{models}) as the ones described in the aforementioned papers by \citet{Vazdekis03, Vazdekis10, Vazdekis12}. As described there, we followed an approach which is as empirical as possible. Hence, contrary to other authors, our SSP models are -- apart from two gap regions, see Section \ref{Preparing the stellar library} --  completely based on an empirical stellar library. In addition, we also made use of extensive photometric libraries rather than theoretical stellar atmospheres in order to transform the theoretical parameters of the isochrones to the observational plane. Furthermore, we did not enlarge our empirical stellar library by including artificial stars. Our models focus on the spectral range between ${\rm 2.5 \, \mu}$m and ${\rm 5 \, \mu}$m in which the \textit{Spitzer} [3.6]$-$[4.5] color can be found.
However, in a forthcoming paper we will combine our models with the already existing ones of \citep{Vazdekis12}. This will enable us to provide the first empirical homogeneous stellar population synthesis models ranging from the optical to the IR.\looseness-2

The paper is structured as follows. In Section 2, we present the IRTF library upon which our SSP models are based. After that, we describe all the steps which had to be carried out in order to prepare the IRTF stars for its use in SSP modelling. We had to test whether the library stars had been correctly flux-calibrated (see Section 3), to find out about the spectral resolution of the library (Section 4) and to determine as accurately as possible the stellar atmospheric parameters of all of the stars (see Section 5). In addition, we checked all of the library stars for peculiarities (see Section 6) before finally correcting the chosen stars by fitting gaps occurring due to telluric absorption lines and extrapolating them to a common dispersion over all the wavelength range (see Section 7). In Section 8 we explain the modelling process, the main ingredients used and in which parameter range our models can be safely applied. We study the behaviour of colors measured from our model spectra as a function of age and metallicity and compare them to photometric predictions from MILES \citep{Vazdekis10} (see Section 9). 
In Section 10, we compare our SSP models to other models available from the literature. Section 11 is dedicated to the discussion of mass-to-light (\textit{M/L}) ratios measured from our models and to the investigation of the dependence of our models on the IMF. Finally, we compare the colors deduced from our SSP models to observed colours of globular clusters and of early-type galaxies (Section 12) and end the paper with a discussion and some concluding remarks (Section 13).

\section{The IRTF library}\label{IRTF}

The IRTF spectral library \citep{Cushing05, Rayner09} contains the spectra of 210 cool stars in the NIR and IR wavelength ranges. All of the stars were at least observed in the spectral range from 0.8 - ${\rm 2.4 \,\mu}$m, but a significant fraction extends up to ${\rm 4 \,\mu}$m, and about half of the stars even up to ${\rm 5 \,\mu}$m.   

The stars were observed with the medium-resolution IR spectrograph SpeX \citep{Rayner03} at the ${\rm3.0\, m}$ IRTF  on Mauna Kea, Hawaii, with a resolving power of $R= \lambda/\Delta\lambda \approx \,$2000. The total wavelength range up to ${\rm 5 \, \mu}$m was covered in only two cross-dispersed instrumental settings. The signal-to-noise ratio (S/N) is better than $\approx \,$100 except for regions of poor atmospheric transmission and for $\lambda  > {\rm 4 \,\mu}$m and all stars have been observed at an airmass $<2$ for good telluric corrections. \looseness-2

The observed sample is composed of stars of spectral types F, G, K, M, some AGB-, carbon- and S-stars, of luminosity classes I-V. Since it was aimed to achieve a high S/N-ratio also in the IR (2.4 - ${\rm5 \, \mu}$m), the selected stars are all relatively bright and situated in the Solar neighbourhood. Consequently, most of the stars of the library are of Solar composition. Unfortunately, this caveat will prevent us from constructing SSP models of highly sub-and supersolar metallicities (see Section \ref{models}).

\section{Testing the flux calibration of the library stars}

Since an important application of our SSP models will be to extract colours from the calculated model spectra and compare them to observations, we have to assess the flux calibration of the stellar spectra. Thus, when integrating the stellar spectra, each star needs to have the correct flux at a given wavelength -- at least with respect to the fluxes at other wavelengths. Modelling requires a correct relative flux calibration, but absolute flux calibration is not a necessary prerequisite. According to \citet{Rayner09}, the IRTF spectra were relatively flux-calibrated and at the same time corrected for absorption due to the Earth's atmosphere by dividing them by a telluric correction spectrum constructed from a theoretical model spectrum of Vega scaled to the spectrum of an observed nearby standard star. The absolute flux-calibration was done by using the Two Micron All Sky Survey \citep[2MASS,][]{Skrutskie00} \textit{JHK}$_{\text{s}}$ photometry.

To test the flux calibration quality of the IRTF spectra carried out by \citet{Rayner09}, we calculated the 2MASS-colors ($J-H$), ($J-K_{\text{s}}$) and ($H-K_{\text{s}}$) from the spectra by integrating over the spectral flux in the 2MASS filters and using a Vega spectrum of \citet{Castelli94} as reference. Then we plotted our calculated colours versus the respective colours extracted from the 2MASS catalogue. In general, we observe quite a big scatter around the identity line, particularly for ($H-K$).

\begin{figure}
\begin{center}
\resizebox{\hsize}{!}{\includegraphics{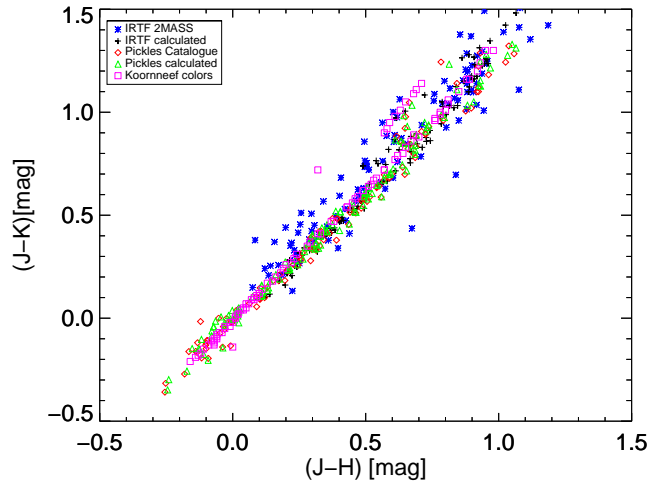}}
\caption{($J-K$) versus ($J-H$) colour-colour plot. Black crosses indicate colours which we calculated from the IRTF spectra, blue stars stand for colours of the IRTF stars extracted from the 2MASS catalogue, whereas red diamonds delineate the colours of the stars of \citet{Pickles98} as catalogued there. Colours which we obtained directly from the spectra of \citet{Pickles98} can be identified by green triangles and magenta squares illustrate the corresponding \textit{JHK}-colours of \citet{Koornneef83}.}
\label{JHK_colors_calculated_catalogue_IRTF_Pickles_Koornneef}
\end{center}
\end{figure}

As a next step, we created a colour-colour plot of ($J-K$) versus ($J-H$) (see Figure \ref{JHK_colors_calculated_catalogue_IRTF_Pickles_Koornneef}). In this plot, we did not only display the respective colours for the IRTF stars as calculated from their spectra and extracted from the 2MASS catalogue, but we also included the colours of the stars of the library by \citet{Pickles98}, once obtained by us from their spectra and again catalogued by the latter, and additionally the \textit{JHK}-colours derived by \citet{Koornneef83}. Figure \ref{JHK_colors_calculated_catalogue_IRTF_Pickles_Koornneef} reveals that the colours of all these samples agree well amongst each other, apart from a significant number of colour values for the IRTF stars originating from the 2MASS catalogue. Consequently, we conclude that in many cases the colours for the IRTF stars deduced from the 2MASS catalogue are somewhat questionable since they seem to be inaccurately determined. This picture of 2MASS photometry is confirmed by \citet{Turner10, Turner11} who claim that it is of uneven quality and low intrinsic precision with errors larger than the cited values for most of the stars in the Galactic plane. However, the fact that the colours which we determined directly by integrating over the IRTF spectra are in good agreement with the corresponding colours from other sources \citep[e.g.,][]{Koornneef83, Pickles98} proves that the flux calibration has been done correctly for the stars of the IRTF library. This is obviously the most important result of our testing.

\section{Determining the spectral resolution}

It is important to find out the spectral resolution of the IRTF stars since this will be also the one of our SSP models. Furthermore, we need to know how the spectra have to be treated during the modelling process. We have to determine the resolution as a function of the wavelength $\lambda$. Equation \ref{spectral_resolution} shows the relation between the respective parameters:

\begin{equation}
 \frac{\text{FWHM}}{\lambda}=\frac{\sigma \cdot 2.35}{c}=\frac{1}{R}
\label{spectral_resolution}
\end{equation}

\begin{figure}
\begin{center}
\resizebox{\hsize}{!}{\includegraphics{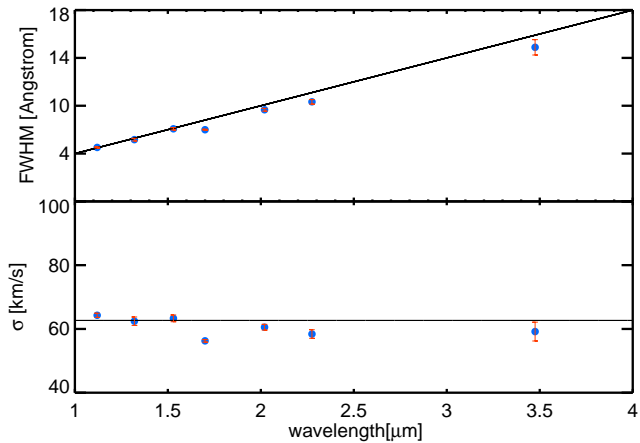}}
\caption{Upper panel: Dependence of the FWHM on wavelength. Shown are the average values for the respective wavelength bins obtained from the sample of 18 test stars of temperatures between 5200 and ${\rm 6000 \, K}$. The red errorbars indicate the errors of the mean values. The black line indicates the FWHM corresponding to a resolving power of $R={\rm2000}$ over this wavelength range. Lower panel: Dependence of the velocity dispersion $\sigma$ on wavelength. The black line delineates the best fit to the data points.}\qquad
\label{Kinematics}
\end{center}
\end{figure}

In Equation (\ref{spectral_resolution}), $R$ denotes the resolving power which, according to \citet{Rayner09}, is $\approx \,$2000. 
We determined the spectral resolution by fitting a combination of spectra of the most recent release of the theoretical Phoenix library \citep{Allard12} to the IRTF spectra. This library comprises spectra of a large parameter range, but we only used the model spectra extending from $\log(g)=-0.5$ to $\log(g)=5$ and $T_{\text{eff}}={\rm 5200 \,K}$ to ${\rm 6000 \, K}$ since we carried out this fitting only for IRTF stars within this temperature range. We finally recovered the FWHM and the radial velocity of our set of test stars by applying the penalized pixel fitting method \citep[ppxf,][]{Cappellari04} following the approach of \citet{Falcon11a}. The spectral fitting was done separately for seven spectral ranges of always around ${\rm 2000 \,\AA}$ width covering the range of the test spectra until the end of the $K$ band. Since the theoretical spectra of the Phoenix library do not model very accurately the spectral features beyond the $K$ band, we just give a single resolution estimate for the reddest spectral range.  
Finally, we calculated one average value for all of our eight wavelength bins from the 18 test spectra which we selected. The upper panel of Figure \ref{Kinematics} shows that the FWHM increases with wavelength from around ${\rm 5 \,\AA}$ in the $J$ band to around ${\rm 10 \,\AA}$ in the $K$ band. Moreover, it looks like this trend continues beyond the $K$ band. However, we should be careful drawing such a conclusion given the fact that we were only able to obtain one further value for the whole wavelength range between the end of the $K$ band and ${\rm 5\,\mu}$m. 

As expected from Equation \ref{spectral_resolution}, the lower panel of Figure \ref{Kinematics} shows that $\sigma$ is virtually constant throughout the whole spectral range at a value of $\approx \,{\rm 60 \, km/s}$. A fit to our data furnishes the exact value of $\sigma=({\rm 60.6 \pm 1.1) \, km \, s^{-1}}$ (see Figure \ref{Kinematics}). Such a value is appropriate to study most galaxies.

\section{Determination of the stellar atmospheric parameters}\label{SAPs}

To use the IRTF stars in SSP modelling it is crucial to determine their stellar atmospheric parameters effective temperature $T_{\text{eff}}$, surface gravity $\log(g)$ and metallicity [Fe/H] as accurately as possible. To measure these parameters, we applied two different methods. The first one consists of compiling the values from the literature and subsequently adjusting them to a master template following the approach carried out by \citet{Cenarro07} (see Subsection \ref{SAPs_literature}). As a second method, we adopted a number of colour-temperature relations from \citet{Alonso96, Alonso99} and from \citet{Conroy12} as well as the relation between temperature, bolometric magnitude, mass and surface gravity from \citet{Straizys81} (see Subsection \ref{SAPs_color_temperature_relations}). The relations by \citet{Alonso96, Alonso99} are the same ones which we later employ in our modelling procedure to transform the theoretical parameters of the isochrones to fluxes and colours (see Section \ref{model_ingredients}). Finally, we checked whether our stellar atmospheric parameters obtained with these two methods were in accordance with the ones expected for a star of the respective spectral type and luminosity class according to the catalogue by \citet{Pickles98}. 
Moreover, we also cross-checked our final set of derived atmospheric parameters with the sample compiled by \citet{Cesetti13} from the literature for many of the IRTF stars.

\subsection{Compiling the stellar atmospheric parameters from literature}\label{SAPs_literature}

\begin{table*}
\caption{ Literature sources containing stellar atmospheric parameters of stars in common with the MILES and IRTF libraries.}
\label{Cenarro_Table}
\centering
\begin{tabular}{l c c}

\hline
Source  & stars in common  &  stars in common \\
 & with the MILES library & with the IRTF library\\
\hline
  IRTF library & 47  & 180 \\ 
  \citet{Prugniel07} & 302 & 34 \\
  \citet{Wu11}  & 373 & 41 \\
  \citet{Cenarro01} [$T_{\text{eff}}$, $\log(g)$] & 348 & 45 \\
  \citet{Cenarro01} [Fe/H] & 308 & 36 \\
  \citet{Gray01} & 43 & 31 \\
  \citet{Valenti05} & 64 & 16 \\
  \citet{Hekker07} & 72 & 11 \\
  \citet{Takeda07} & 90 & 17 \\
  \citet{Mishenina04, Mishenina06, Mishenina08} & 149 & 15 \\
  \citet{Kovtyukh03, Kovtyukh06} [$T_{\text{eff}}$] & 92 & 16 \\
  \citet{Kovtyukh08} [$\log(g)$, [Fe/H]] & 27 & 6 \\
  \citet{Luck07} & 53 & 4 \\
  \citet{Santos04, Santos05} & 35 & 8 \\
  \citet{Sousa08, Sousa11} & 37 & 5 \\
  \citet{Kovtyukh07} & 11 & 10 \\
  \citet{Lyubimkov10} & 5 & 6 \\
  \citet{Edvardsson93} & 94 & 11 \\
  \citet{Borkova05} & 210 & 18 \\
  \citet{Thevenin98} & 273 & 31 \\
  
\hline
\end{tabular}
\end{table*} 

We started by querying the literature in order to find all relevant sources containing spectroscopically determined stellar atmospheric parameters. Then, we checked whether any of the 180 stars that we used from the IRTF library were contained in the different references. Whenever this was the case, we matched the respective sample of stars from the literature with the 985 stars of the MILES stellar library \citep{Sanchez06} to find stars in common. In order to be fully consistent with our model predictions in the optical range, the stellar parameters of the MILES stars as determined by \citet{Cenarro07} served as the basic reference sample against which all the other references were calibrated. Finally, we ended up with 17 sources from the literature which have stars in common with both the MILES and the IRTF stellar libraries (see Table \ref{Cenarro_Table}).


  

After that, the three parameters $T_{\text{eff}}$, $\log(g)$ and [Fe/H] of the stars in common between one literature source and MILES were separately compared to each other. For all three parameters, we subsequently obtained the best-fitting lines mapping the relations between the MILES values and their equivalents from the respective literature reference. When carrying out these fits, we first used all stars in common between MILES and the literature source. In the next step, we discarded all those stars showing a value $\Delta{\rm (parameter) = \lvert parameter_{MILES} - parameter_{reference}\rvert}$ larger than ${\rm1\,\sigma}$ (i.e., 1-$\sigma$ clipping) and performed another fit. The offset parameter A and regression coefficient B of this latter fit were then used to correct the stellar parameters of all the stars of the respective reference. Like this, we also enlarged significantly the initial reference sample which consisted of MILES stars only. Consequently, also those references \citep{Santos04, Santos05, Sousa08, Sousa11, Kovtyukh07, Kovtyukh08, Lyubimkov10} which a priori did not exhibit enough stars in common with the MILES library to obtain reliable fits could get adjusted to the MILES scale. According to \citet{Cenarro07}, for this a minimum number of 25 stars in common is needed. 

\begin{figure}
\begin{center}
 \resizebox{\hsize}{!}{\includegraphics{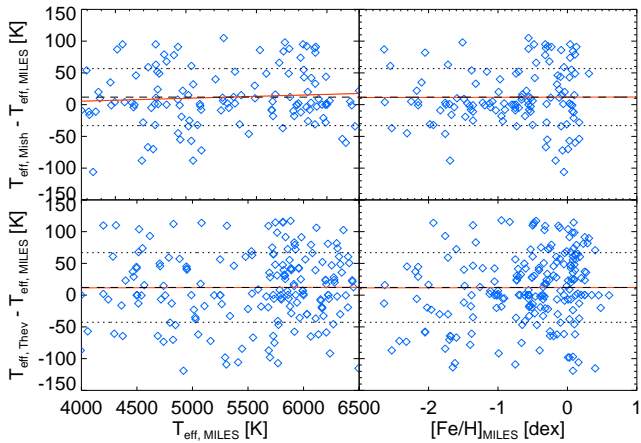}}
  \caption{Differences in temperature between MILES and two literature sources as a function of temperature (left-hand panels) and metallicity (right-hand panels), respectively. These two sources are the catalogue of \citet{Mishenina04, Mishenina06, Mishenina08} (upper panels) and the one of \citet{Thevenin98} (lower panels). The red solid lines are the ones indicating the best fits, the black dashed lines represent the mean temperature differences between the two samples and MILES, whereas the black dotted lines show the one $\sigma$ standard deviations from these mean differences in temperature.}\qquad
  \label{Delta_Teff_versus_Teff}
\end{center}
\end{figure}

For all of our references, we also created plots showing the differences between the respective parameter values and the MILES ones versus the MILES values of all three different parameters. In the case of most of our references, the $\sigma$ standard deviations of the parameter differences do not exceed ${\rm75 \, K}$ in $T_\text{eff}$, 0.2 in $\log(g)$ and 0.05 in [Fe/H] (see Figure \ref{Delta_Teff_versus_Teff}).

Finally, we found that the parameters of 44 of our 180 IRTF stars could be directly adopted from MILES. 32 further stars were contained in at least one reference, whereas 20 additional ones were discovered in several references. The final parameters $p$ of the latter ones were calculated applying the following equation, given by \citet{Cenarro07}: 

\begin{equation}
 p = \frac{\sum\limits_{i=1}^N p_i^*/\sigma_i^2}{\sum\limits_{i=1}^N 1/\sigma_i^2} \,.
\label{Equation_multiple_sources}{SAPs_color_temperature_relations}
\end{equation}

\noindent{Here, p$^*_i$ denotes the parameter of one literature reference corrected to the MILES scale, and $\sigma_i$ corresponds to the standard deviation of this reference with respect to the MILES system.}



We were not able to derive stellar atmospheric parameters for 79 of our stars from the literature. For these stars, colour-temperature relations had to be applied (see next Section \ref{SAPs_color_temperature_relations}).

\subsection{Applying colour-temperature relations}\label{SAPs_color_temperature_relations}

Since literature values are not available for all of our used IRTF stars, we also followed a second, different approach to determine the effective temperatures of the stars of the IRTF library. Therefore, we used the colour-temperature relations determined by \citet{Alonso96} for dwarfs and for giants \citep{Alonso99} of spectral types F0-K5 for colours ranging from optical to the NIR. \citet{Alonso96, Alonso99} indicated in which colour ranges their relations are applicable. These are metallicity-dependent relations, although for the ($J-K$) colour the sensitivity to this parameter is negligible. We consequently calculated the $J$ and $K$ magnitudes from the spectra of all of our stars by convolving them with the respective filter response functions. For the stars of the IRTF library which only extend up to ${\rm 2.4 \,\mu m}$ we had to use the slightly metal-dependent relation for the ($J-H$) colour in order to calculate the temperatures, since their spectra do not cover the whole spectral range of the $K$ filter. It should be noted that the $J$, $H$ and $K$ filters used by \citet{Alonso96, Alonso99} for their observations at the Carlos Sanchez Telescope (TCS) at the Observatorio del Teide in Tenerife are non-standard ones. The filter response functions are given in the appendix of \citet{Alonso94}. 

The relations of \citet{Alonso96, Alonso99} are however not valid for M-stars. For them, we adopted the temperature values calculated by \citet{Conroy12} for these stars using the temperature-colour-relations by \citet{Ridgway80} and \citet{Perrin98} for M giants and by \citet{Casagrande08, Casagrande10} for M dwarfs.

\begin{figure}
\begin{center}
 \resizebox{\hsize}{!}{\includegraphics{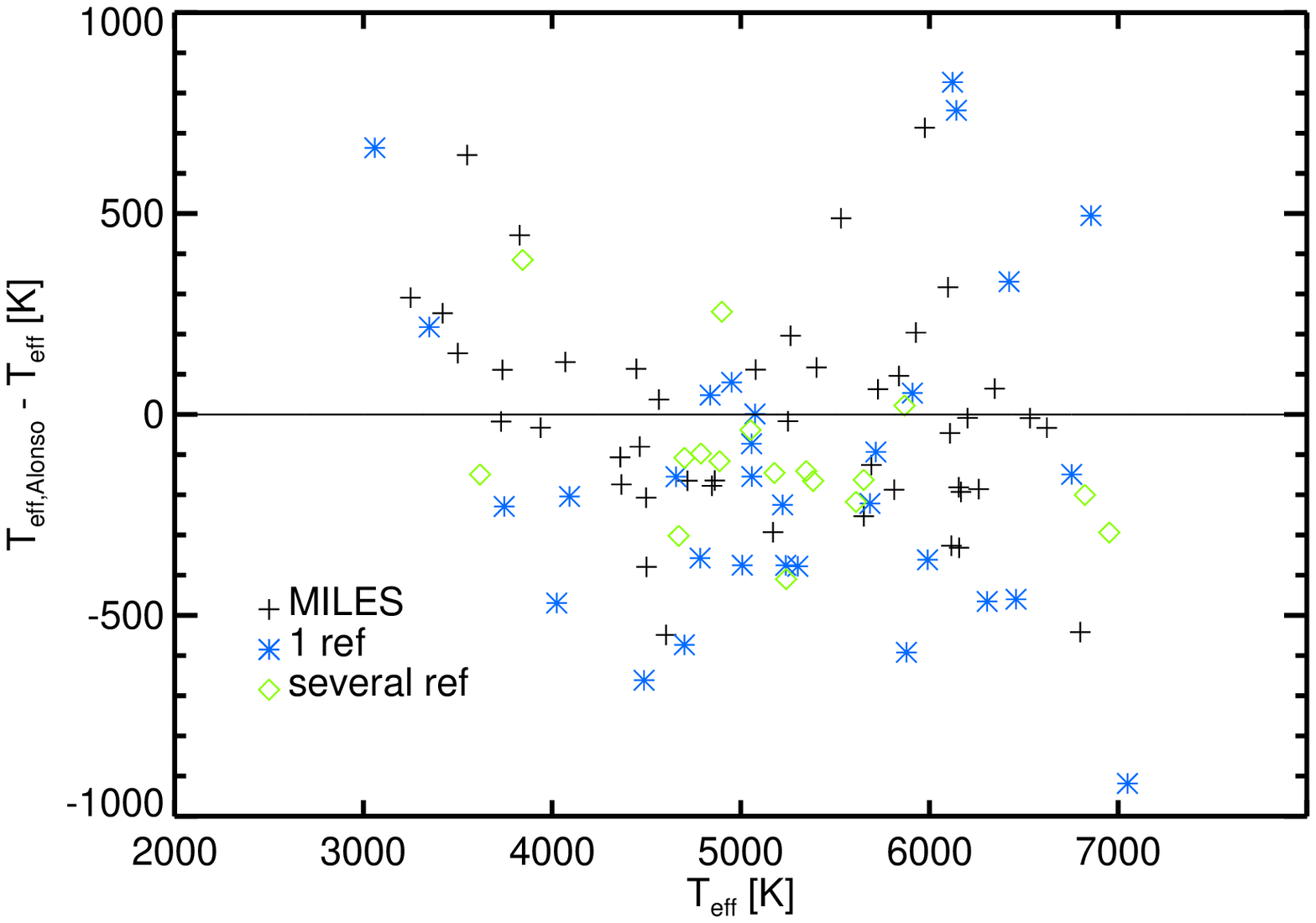}}
  \caption{Differences between effective temperatures determined following \citet{Alonso96, Alonso99, Conroy12} and according to a literature compilation \citep{Cenarro07} versus the latter ones. The black pluses indicate $T_\text{eff}$ values of the respective stars directly from the MILES catalogue \citep{Cenarro07}, whereas the blue asterics and green diamonds indicate that the corresponding $T_\text{eff}$ is based on one or several literature sources, respectively. The solid black line marks the identity between the effective temperatures determined using \citet{Alonso96, Alonso99} and \citet{Cenarro07}.}\qquad
  \label{Teff_comparison_Benny_Alonso_refs}
\end{center}
\end{figure}

Figure \ref{Teff_comparison_Benny_Alonso_refs} compares the temperatures calculated using the aforementioned relations of \citet{Alonso96, Alonso99} and \citet{Conroy12} to the temperatures determined beforehand by means of the approach of \citet{Cenarro07} for those stars with atmospheric parameters in the literature.

The variations are reasonable in general, although they are quite pronounced for some individual stars and the temperatures from the color-temperature relations appear to be a bit lower on average than the ones from the literature compilation. Consequently, in order to decide which of the two temperature values to adopt we performed a cross-check with the catalogue by \citet{Pickles98}. This library consists of 131 stellar spectra extending from ${\rm1150 \,\AA}$ to ${\rm10620 \,\AA}$ (half of them even further until ${\rm25000 \,\AA}$) which were artificially compiled based on several sources. We searched for counterparts having the same spectral types like our IRTF stars among the Pickles stars. Before, we had a detailed look at all the spectra by eye in order to make sure that their spectral types were properly determined. Then we checked what temperature a star of a specific spectral type should have according to \citet{Pickles98}, in order to get an idea about which of our two temperature values is most likely the more reliable one. In the end, we substituted the temperature values from our \citet{Cenarro07} approach by the ones calculated using the colour-temperature(-metallicity) relations of \citet{Alonso96,Alonso99} for a number of stars. For most of these stars, only one literature reference, different from MILES, had been found. For a small number of stars, we used the fast approach of taking directly the temperature values given in the Pickles-library for a star of a similar spectral type. 

For the five carbon stars in our sample, we adopted the effective temperature values obtained by \citet{Bergeat01} as the arithmetic mean of the temperatures estimated from the integrated flux and from calibrated colour indices.

We also calculated $\log(g)$ for all IRTF stars for which this was feasible due to their spectral type and luminosity class by applying the relation between surface gravity, temperature, stellar mass and bolometric magnitude of \citet{Straizys81}. For those stars for which this calculation yielded a value of the surface gravity which is in better agreement with the value expected from the luminosity class given in the IRTF library, we substituted the value. 

\begin{figure}
\begin{center}
\resizebox{\hsize}{!}{\includegraphics{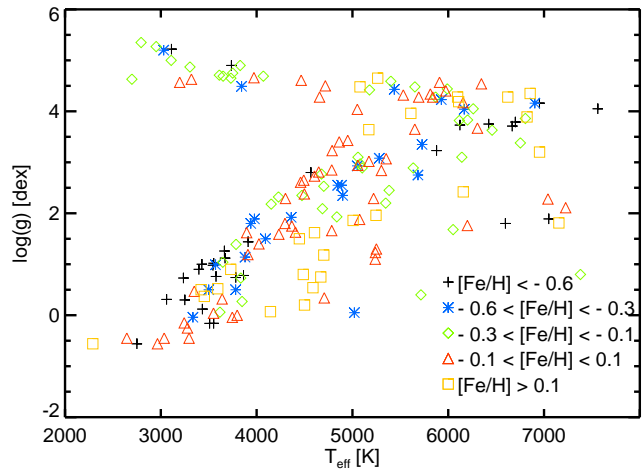}}
  \caption{Distribution of our IRTF stars in the $\log(g)$-$T_{\text{eff}}$ parameter space. Blue pluses indicate stars of a metallicity lower than ${\rm[Fe/H]=-0.6}$, orange stars stand for stars in the metallicity range between ${\rm[Fe/H]=-0.6 \, dex}$ and ${\rm[Fe/H]=-0.3}$, whereas red diamonds delineate stars of metallicities between ${\rm[Fe/H]=-0.3}$ and ${\rm[Fe/H]=-0.1}$. Stars of around solar metallicities (${\rm-0.1 \,<\, [Fe/H] \,<\, 0.1}$) can be identified by green triangles and black squares illustrate all stars of metallicities greater than ${\rm[Fe/H]=0.1}$.} \qquad
  \label{SAP_distribution_logg_versus_Teff_final}
\end{center}
\end{figure}

Table \ref{Table_final_SAPs} in the appendix contains our final set of stellar atmospheric parameters as used later in the SSP modelling (see Section \ref{models}). 
In this Table, we also indicate how the parameters were determined. In general, the values from our \citet{Cenarro07} approach were always preferred over the ones from the colour-temperature relations as long as they were in reasonable agreement with the expected values from \citet{Pickles98}. Figure \ref{SAP_distribution_logg_versus_Teff_final} illustrates the distribution of these parameters in the $\log(g)$-$T_{\text{eff}}$ plane. The respective stars are colour-coded according to their metallicities. From Figure \ref{SAP_distribution_logg_versus_Teff_final}, it is obvious that a good parameter coverage in $T_{\text{eff}}$ and $\log(g)$ only exists for stars of around Solar and for slightly sub- and supersolar metallicities. Subsequently, we model SSPs just within this metallicity range (see Section \ref{models}).

\section{Stellar content of our models} 

In this Section, we discuss some aspects of the final sample of stars on which our SSP models are based. At first, we comment on the stars which we finally had to discard or at least reduce in weight due to various reasons (see Subsection \ref{peculiar_stars}). Since the SSP spectra in the NIR and IR depend mostly on the coolest stars, in Subsection \ref{Cool_stars}, we describe what kind of cool stars entered the models. Moreover, we give the fractions of the final modelled bolometric flux due to a particular subsample of cool stars.

\subsection{Checking for peculiar stars}\label{peculiar_stars}

For their later use in SSP modelling, it is very important to make sure that we only consider spectra of normal stars -- i.e., stars without any peculiar spectral features such as very strong emission lines. In order to decide which stars to discard from our sample of IRTF stars or of which we should at least reduce the weight with which they will contribute to the modelling process, we took advantage of the so-called interpolator \citep[see Section \ref{model_computation}; for a detailed explanation see the appendix of][]{Vazdekis03}. 

\begin{figure}
\begin{center}
 \resizebox{\hsize}{!}{\includegraphics{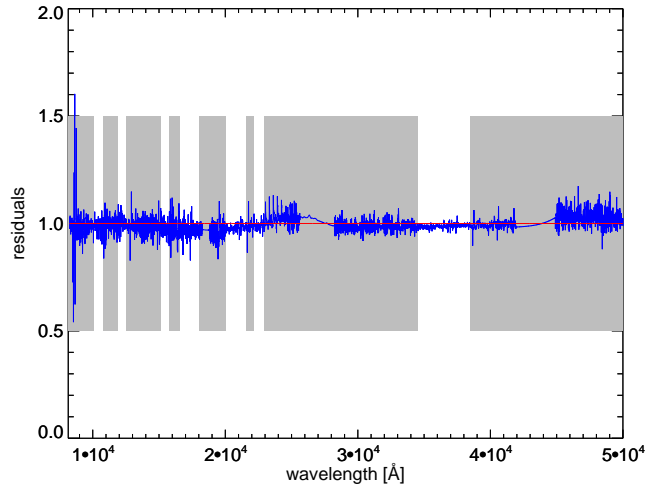}}
  \caption{Example of a residual plot. We divided the spectrum of HD 21018 ($T_{\text{eff}}=5353 \,\text{K}$, $\log(g)=3.07$, ${\rm[Fe/H]=0}$) by the one of its artificially created counterpart obtained by using the interpolator. The red line runs parallel to the $x$-axis at $y=1$. The grey-shaded areas are the parts of the spectrum which were masked out when getting the statistics (mean value, standard deviation, maximum and minimum value) of the plot, since they are more affected by noise and telluric features and thus less reliable for these purposes (see text).}\qquad
  \label{IRTF25_overplotted_int_residuals_masked}
\end{center}
\end{figure}

We then used this programme in order to create artificial stars with the same parameters as all the stars of the IRTF library. When carrying out this interpolation, we always excluded the actual star whose spectrum should be reproduced artificially from the list of stars contributing to the interpolation. Then, we created plots showing the residuals between one spectrum and its artificial counterpart. Before analysing these residual plots statistically, we masked out some wavelength regions hampered by strong telluric features and spikes (see Figure \ref{IRTF25_overplotted_int_residuals_masked}).
Stars showing a standard deviation smaller than around 0.03 and a maximum and minimum value deviating by no more than around 20 per cent from unity qualified to contribute with their full weight to our models. We had a detailed look at all those stars which exceeded these limits. For many of these stars, the interpolator was simply unable to create a reliable counterpart of the same parameters due to the limited number of stars in the respective regions of the parameter space. This was in particularly the case for stars of extremely high and low metallicity, C-stars, S-stars, dwarfs of very low temperature, dwarves of high temperature and comparably low surface gravity, high-temperature giant stars and low-temperature supergiants. Whenever it was feasible, we rather reduced the weights with which these stars contribute to the SSP models than discarded them completely. The exceptions were star HD 142143) which features a very bad residual plot with a never constant baseline, stars HD 14386 and HD 69243 because of their multiple emission lines, and stars GJ 1111 and HD 6474 for which wrong temperature values were clearly assumed. Moreover, the four S-stars (BD+442267, HD 64332, HD 44544, HD 62164) had to be eliminated, as they are long-period variables and also contain some peculiar spectroscopic features. We reduced the weight of 18 stars to 50 per cent of the weight given to a good star (HD 40535, HD 87822, HD 17918, HD 126660, HD 202314, HD 179870, HD 212466, BD+60265, HD 339034, HD 39801, HD 206936, HD 95735, HD 14488, Gl 268, HD 196610, HD 42581, HD 35661, HD 14404) and of one star (Gl 406) to only 25 per cent of the weight of a good star, and excluded a further nine stars completely (see above).

\subsection{Cool stars}\label{Cool_stars}

Figure \ref{SAP_distribution_logg_versus_Teff_final} in Section \ref{SAPs_color_temperature_relations} reveals that despite the generally poor coverage of the stellar atmospheric parameter space, potentially very important and relevant stars for SSP modelling in the IR like cool dwarfs and supergiants form an integral part of our finally used stellar sample. Hence, our sample of 180 stars contains in total 32 AGB stars, amongst them 16 very cool supergiants of temperatures between ${\rm T_{eff}=2300 \, K}$ and ${\rm T_{eff}=3500 \, K}$ and 17 cool dwarfs with temperatures ranging from ${\rm T_{eff}=2700 \, K}$ to ${\rm T_{eff}=3840 \, K}$. Table \ref{Cool_stars_names} lists all the mentioned cool dwarfs and supergiants together with their stellar atmospheric parameters. 

Consequently, we are able to reach masses as small as $0.15 M_{\odot}$. In the standard case of a Kroupa-like IMF, the contribution of these cool very low main sequence red dwarfs of masses smaller than $0.5 M_{\odot}$ to the total bolometric flux of one particular SSP-model is between $1 \, \%$ at ages of around 1 - ${\rm 2 \, Gyr}$ and $5-6 \, \%$ at the oldest ages, irrespective of the metallicity. For a more bottom-heavy IMF (see Section \ref{IMF}), these fractions increase to values of between 6 and $14 \, \%$. 
As expected, the cool AGB stars are particularly important at the lower ages between 1 and ${\rm 2.5 \, Gyr}$ \citep[compare also][]{Maraston05, Maraston09}. In this age range, the AGB stars are responsible for more than $20 \, \%$ of the total bolometric flux of the models, whereas this percentage has dropped down to only $\approx 10 \, \%$ at the oldest ages of ${\rm 10 - 12 \, Gyr}$. 
Among the AGB stars of our sample, also some carbon stars can be found. Our input sample of 180 stars contains five carbon stars, stars HD 31996, HD 57160, HD 76221, HD 44984 and HD 92055 of temperatures between 2300 and ${\rm 3300 \, K}$. Carbon stars contribute up to ${\rm 5 \, \%}$ of the total bolometric flux in the case of solar and higher metallicities at the youngest ages covered here (${\rm 1 - 2 \, Gyr}$). This fraction increases for subsolar metallicities to values of ${\rm \approx 10 \, \%}$. For older ages, carbon stars do not play a role and their contribution to the total bolometric flux is negligible. Oxygen-rich stars, however, are not included in the IRTF library or at least it is is not explicitly pointed out. 
All the presented estimates of the percentages of the total bolometric flux assigned to the various categories of cool stars are based on the BaSTI-isochrones of \citet{Pietrinferni04} (more details on the modelling process and on the necessary ingredients in Section \ref{models}). Although it is not stated explicitly in \citet{Pietrinferni04}, they do take the so-called third dredge-up, during which convection transports the carbon produced in the stellar interior by helium fusion up to the stellar surface, into account (Cassisi, private communication and \citet{Vazdekis15}). For doing so, they apply the analytical relationships for the trend of the surface luminosity and effective temperature as a function of the CO core mass provided by \citet{Wagenhuber96, Wagenhuber98} and a correction to the $(J-K)$ colour which depends on the effective temperature from \citet{Bergeat01}.

\begin{table}
\caption{List of the coolest stars of the library: The first column gives the effective temperatures $T_{\text{eff}}$. Columns 2 and 3 contain the surface gravities $\log(g)$ and the metallicities [Fe/H], respectively. The official star name can be found in Column 4 and the type of star (carbon, AGB or M dwarf) is indicated in Column 5.}
\label{Cool_stars_names}
\centering
\begin{tabular}{l r r l l}
 \hline
  
 $T_{\text{eff}}$ [K]& $\log(g)$ [dex]& [Fe/H] [dex] &  star name &type\\

\hline
   2314&      -0.56&       0.260& HD31996 & carbon \\
   2645&      -0.45&      -0.039& HD76221 & carbon  \\
   2698&       4.63&      -0.149& Gl406 & M dwarf \\
    2750&      -0.56&      -2.609& HD207076 & AGB \\
    2794&       5.35&      -0.147& Gl51  & M dwarf \\
    2951&       5.27&      -0.151& Gl866  & M dwarf \\
    2965&       -0.56&     -0.060& HD92055 & carbon \\
   3030&       5.20&      -0.544& Gl213 & M dwarf \\  
   3035&      -0.45&      -0.074& HD44984& carbon   \\
   3060&       0.31&      -1.016& HD196610 & AGB \\
   3108&       5.00&      -0.247& Gl273 & M dwarf \\
  3111&       5.22&      -0.718& Gl299 & M dwarf \\
  3197&       4.57&       0.000& Gl581 & M dwarf \\
  3236&       0.73&      -2.609& HD156014 & AGB \\
  3250&      0.30&      -1.005& HD18191 & AGB \\
  3278&      -0.25&      -0.069& HD339034 & AGB \\
  3295&      -0.45&       0.046& HD57160 & carbon  \\
  3304&       4.87&      -0.158& Gl388 & M dwarf \\
   3319&       4.63&       0.000& Gl381 & M dwarf \\
  3349&       0.47&       0.026& HD40239 & AGB \\
  3397&       0.90&      -1.396& HD214665 & AGB \\
  3420&       0.50&       0.140& HD175865 & AGB \\
  3428&       1.00&      -1.464& HD204585 & AGB \\
  3434&       0.12&      -2.267& HD108849 & AGB \\
  3454&       0.37&       0.183& BD+60265 & AGB \\
  3509&      -0.16&      -0.772& HD14488 & AGB \\
  3544&       0.98&      -0.959& HD19058 & AGB \\
  3550&       0.04&       0.030& HD39801 & AGB \\
   3551&      -0.16&      -0.693& HD14469 & AGB \\
  3571&       0.99&      -0.508& HD4408 & AGB \\
   3576&       0.76&      -2.609& HD94705 & AGB \\
  3594&       0.52&       0.177& HD236697 & AGB \\
  3611&       4.71&      -0.192& Gl806 & M dwarf \\
  3617&       0.06&      -0.201& HD35601 & AGB \\
  3640&       0.32&      -0.070& HD14404 & AGB \\
  3650&       4.69&      -0.149& HD209290 & M dwarf \\
  3730&       0.90&       0.270& HD219734 & AGB \\
  3731&       4.65&      -0.161& HD42581 & M dwarf \\
  3737&       4.90&      -1.500& HD36395 & M dwarf \\
   3746&      -0.04&       0.046& HD206936 & AGB \\
   3750&       4.75&      -0.156& Gl268 & M dwarf \\
  3781&       0.50&      -0.458& HD10465 & AGB \\
  3782&       0.74&      -0.928& HD216946 & AGB \\
  3797&       0.00&       0.064& HD212466 & AGB \\
   3828&       4.90&      -0.200& HD95735 & M dwarf \\
  3843&       4.49&      -0.408& HD201092 & M dwarf \\
  3848&       0.27&      -0.151& HD185622 & AGB \\
  3863&       0.78&      -0.626& HD23475 & AGB \\
\hline
\end{tabular}
\end{table}

\section{Preparing the stellar spectra}\label{Preparing the stellar library}

\begin{figure}
\begin{center}
 \resizebox{\hsize}{!}{\includegraphics{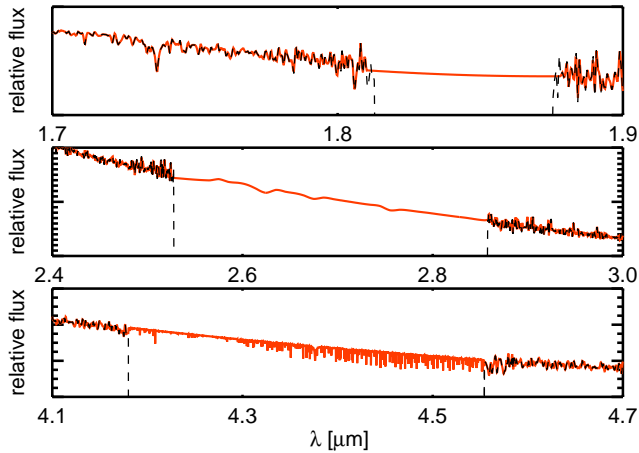}}
  \caption{The three ranges of the spectrum of HD 213306 showing gaps. The original spectrum is delineated with a black dotted line. We overplotted the gap-corrections in red.}\qquad
\label{HD213306}
\end{center}
\end{figure}

In this Section, we will briefly describe the final treatments which we had to carry out on the spectra before being able to use them for SSP modelling. First of all, it turned out that all of the IRTF spectra show three characteristic gaps, one between 1.80 and ${1.87 \,\mu m}$ (i.e., between the $J$ and the $H$ band), another one between $\approx \, 2.50$ and $2.80 \,\mu m$ and a third one between $\approx \, 4.20$ and ${\rm 4.50 \,\mu m}$. The exact limits of the latter two ones vary slightly from spectrum to spectrum. According to \citet{Rayner09}, the first gap is mainly due to the instrumental setting, whereas the second and third ones are caused by strong telluric absorption in the respective wavelength ranges. We interpolated the spectra linearly within the first two gaps using IRAF. Whenever it was feasible, we replaced those two gaps by the theoretical spectra of the BaSeL-library \citep{Lejeune97, Lejeune98, Westera02} enabling us to include also some spectral features. The spectra of this library extend over the large wavelength range ${\rm91 \,\AA}$ to ${\rm 1600000 \, \AA}$, however at a rather low resolution of only 10 - ${\rm20 \, \AA}$, and were compiled for a close grid of stellar atmospheric parameters. The upper two panels of Figure \ref{HD213306} show the spectral ranges around the first two gaps of the star HD 213306 ($T_{\text{eff}}=6051 \,\text{K}$, $\log(g)=1.69$, ${\rm[Fe/H]=-0.12}$) containing also the corrections carried out within these gaps, based on the corresponding BaSeL-spectrum.

There is a strong molecular CO absorption band in the range between ${\rm 4.20 \,\mu m}$ and ${\rm 5 \, \mu m}$ which affects the ${\rm 4.50 \,\mu m}$ filter \citep[e.g.,][]{Monson12, Peletier12, Meidt14}. Unfortunately, this CO absorption falls in the gap present in all of the IRTF spectra in the region between ${\rm 4.2\, \mu m}$ and ${\rm 4.5\, \mu m}$, and thus prevents us from  properly accounting for this flux loss in the ${\rm 4.50\, \mu m}$ band. 
As can be seen from Figure 10 in \citet{Monson12}, this CO absorption affects stars of temperatures cooler than about ${\rm 4500 \,K}$, i.e., the ones which contribute most to our models in the IR \citep[see also][]{Peletier12}.

In order to correct our spectra for the effect of the CO in the ${\rm 4.5 \, \mu m}$ band, we made use of the theoretical Phoenix library \citep{Allard12} for filling in the gap. Thanks to their high resolution of ${\rm 0.1\, \AA}$ in our IR wavelength range, the spectra of this library account for the effect of the CO absorption. In addition, they cover all of our temperature range from ${\rm \approx 2000\, K}$ to ${\rm \approx 7500 \, K}$ with a spacing of ${\rm 100\, K}$, contain surface gravities ranging from ${\rm -0.5\, dex}$ to ${\rm 5.50\, dex}$ in steps of ${\rm 0.5\, dex}$ and are available for metallicities between ${\rm [Fe/H] = -4.0 \, dex}$ and ${\rm [Fe/H] = 0.5 \, dex}$ with ${\rm \Delta = 0.5 \, dex}$. The lower panel of Figure \ref{HD213306} illustrates the fitting of this third gap by the theoretical Phoenix spectrum with the closest agreement in the stellar atmospheric parameters to the ones of the respective IRTF star. Finally, this approach of correcting for the CO absorption feature enables the reader to deduce colours involving the ${\rm4.5 \, \mu m}$ band from our SSP model spectra without having to apply a missing-flux correction accounting for the gap.




We note that the spectra of the IRTF library are not given with a common pixel size, but that it grows from around ${\rm 2 \,\AA \, pixel^{-1}}$ at ${\rm 0.8 \,\mu m}$ to around ${\rm 10 \, \AA \, pixel^{-1}}$ at ${\rm 5 \, \mu m}$. In order to use the spectra for SSP modelling, it is necessary to have them at a common pixel size. We thus rebinned them to ${\rm \Delta \lambda = 2.73 \, \AA}$, since this is the value corresponding to a wavelength of around ${\rm 1 \,\mu m}$.

We still had to extrapolate the spectra of all of our used IRTF stars which do not already intrinsically extend up to ${\rm5 \,\mu m}$ to this value. In order to fulfill this task, we made again use of the interpolator. At first, we created artificial counterparts reaching up to ${\rm5 \, \mu m}$ for all of the IRTF stars whose spectra did not extend beyond ${\rm4 \, \mu m}$. This interpolation was based on all stars with spectra observed up to ${\rm5 \,\mu m}$. Then, we combined the artificially created part of the spectrum between 4 and ${\rm5\, \mu m}$ with the actual spectrum of the respective star reaching up to ${\rm4 \,\mu m}$. Therefore, we selected the last part of the original stellar spectrum from around 39500 to ${\rm40000 \,\AA}$ and calculated the mean offset in this spectral range between the spectrum of the ${\rm4 \, \mu m}$  IRTF-star and the corresponding star of the same parameters and added this offset to the latter one. We proceeded in the same way for the 29 stars whose spectra did not reach beyond ${\rm 2.4 \,\mu m}$. Now we could take advantage of all those stars which we extrapolated from 4 to ${\rm 5 \, \mu m}$ before in order to enlarge our input sample of stars for performing the interpolation. Like this, we ended up with spectra extending up to ${\rm 5 \, \mu m}$ for all IRTF stars. 

In the first part of Table \ref{Parameter_coverage} we present the spectral properties of the 180 spectra of our final stellar library which we will subsequently use in our SSP models (see next Section).

\section{Infrared SSP models}\label{models}

SSP models are a very important tool for studying clusters and galaxies and in particular to determine their ages and metallicities. These comparisons can be carried out by means of full spectrum fitting \citep[see, e.g.,][]{Koleva11} or by using information about integrated colours or line strength indices. 

In the optical wavelength regime, quite a number of SSP models are readily available \citep[e.g.,][]{Vazdekis99, Bruzual03, Conroy10, Vazdekis10, Maraston11, Conroy12, Vazdekis12}, whereas in the NIR and IR wavelength ranges, their number is more limited. This is particularly true for the IR. Whereas examples for the NIR include for example \citet{Mouhcine02, Maraston09}, there do not exist any SSP models beyond the $K$ band which are mainly based on empirical stellar spectra and which predict model spectra at moderately high resolution.

\subsection{Main ingredients}\label{model_ingredients}

We calculated our SSP models based on the isochrones of \citet{Girardi00} (hereafter: Padova00-isochrones) and also the ones of \citet{Pietrinferni04} (BaSTI-isochrones). Both sets of model isochrones cover a wide range of ages and metallicities - in fact much wider than needed for our purposes given the limits in stellar atmospheric parameter coverage of the IRTF library (see Section \ref{SAPs}). Consequently, we calculated SSP models only for the five metallicities ${\rm [Fe/H] = -0.35, -0.25, 0.06, 0.15, 0.26}$ in the case of BaSTI and ${\rm[Fe/H]=-0.40, 0, 0.22}$ for Padova00. Please note that the BaSTI isochrone corresponding to a metallicity of ${\rm[Fe/H]=0.15}$ is not yet publicly available since it was calculated exclusively for this work (and others) by S. Cassisi \citep[see also][] {Vazdekis15}.

We also set some limits on the age range for which we obtained SSP models. So, we did not determine SSP models for ages younger than 1 Gyr, since at these ages, the contribution of young hot stars of $T_{\text{eff}}>7000 \, {\rm K}$ becomes important. Unfortunately, the IRTF library is in general poor in these stars. The same holds for AGB stars which dominate the IR light at ages between 1 and 2 Gyr. In addition, this stellar evolutionary phase is difficult to model. Subsequently, predictions differ a lot from each other depending on the synthetic prescriptions used by the authors. This can be seen by comparing, e.g., the isochrone set of \citet{Marigo08} to the one of \citet{Girardi00}. Apart from that, the IRTF library is also in general poor in luminous giants (AGB stars, see Section \ref{SAPs}), leading to a significant decrease in quality for SSP models of younger ages. 

We took various IMFs into account, obtaining models for IMFs of single slopes between 0.3 and 3.3. These slopes were either kept constant over the whole mass range or a shallower slope was adopted for masses smaller than $0.6 \, M_{\odot}$ \citep[so called unimodal and bimodal IMFs, respectively,][]{Vazdekis96}. In bimodal IMFs, the contribution of low mass stars is significantly reduced. We also calculated SSP models based on the Kroupa and revised-Kroupa IMFs \citep{Kroupa01}. In order to keep the following discussion more clear, we concentrate on the SSP models and their predictions based on the Kroupa IMF. This is justified since the influence of the IMF on our models is almost negligible. Yet, possible impacts of different IMFs will be discussed in Section \ref{IMF}. 

\subsection{Computation of the models}\label{model_computation}

We present here an extension of our models \citep{Vazdekis03, Vazdekis10, Vazdekis12}, which are so far available for the optical wavelength range between ${\rm3465 \,\AA}$ and ${\rm 9469 \,\AA}$,  to the IR, out to ${\rm5 \mu m}$. The general modelling procedure follows very closely the approach carried out before in the optical. 
The general idea of this approach is to populate isochrones of various ages and metallicities with the spectra of our input stellar library (see Section \ref{IRTF}) according to the prescription given by a chosen IMF. Consequently, if the isochrone demands a star of a particular set of stellar parameters $T_{\text{eff}}$, $\log(g)$ and [Fe/H], we try to reproduce this star as accurately as possible. For this, we use a so-called interpolator which creates a star of the desired parameters based on the input stellar library. 

The interpolator is a code which calculates a stellar spectrum for any given set of stellar atmospheric parameters, i.e., effective temperature, surface gravity and metallicity, based on the stellar spectra of a whole library of input stars. The general idea of the interpolator code is to search for stars which are contained within up to eight cubes in the stellar atmospheric parameter space. One corner of these cubes is always centred on the parameters of the requested star. Depending on how densely the parameter space is populated with stars in a certain region, the size of these cubes varies between a minimum and a maximum value. We adjust these input values of the interpolator to the scale required by the IRTF library. The parameters temperature and surface gravity are always taken into account, however, the metallicity only for stars of temperatures between 3250 and ${\rm7000 \,K}$. For hotter and cooler stars, respectively, we neglect this parameter because the IRTF library does not contain a large enough number of stars in these temperature ranges in order to cover the parameter space in metallicity in a sufficient way (see Figure \ref{SAP_distribution_logg_versus_Teff_final} and Section \ref{SAPs_color_temperature_relations}).

After calculating the stellar spectrum of the requested parameters, each of these spectra is normalised by convolution with the $K$ band filter and the photometric libraries and colour-temperature relations of, e.g., \citet{Alonso96, Alonso99} are used to convert the theoretical parameters ($T_{\text{eff}}$, $\log(g)$, [Fe/H]) to the observational plane. We chose to normalise our stellar input spectra to the $K$-band, since this band is located well in the middle of the spectral range covered by the IRTF-library and spectrally close to the IR wavelength range between 2.5 and ${\rm 5 \, \mu m}$ on which we focus in this paper. Finally, the stellar spectra are integrated along the isochrones. 

The procedure of calculating a SSP spectral energy distribution (SED) is summarised as:

\begin{equation}
\begin{split}
 S_{\lambda}(t, \text{[Fe/H]}) = \int_{m_1}^{m_t} S_{\lambda}(m,t,\text{[Fe/H]})\cdot\\
 N(\text{IMF},m,t)\cdot F_K(m,t,\text{[Fe/H]})\mathrm{d}m,
\label{SSP_SED}
\end{split}
\end{equation}

\noindent{where $S_{\lambda}(m,t,\text{[Fe/H]})$ is the spectrum of a single star of mass $m$ and a certain lifetime and metallicity which are determined by the adopted isochrone.} $N$(IMF,$m$,$t$) is the number of this kind of star which depends on the chosen IMF. The smallest and largest masses which exist for a single stellar population of a given age are designated by $m_1$ and $m_t$. $F_K(m,t,\text{[Fe/H]})$ describes the flux of each single stellar spectrum in the $K$ band. We stress again that the fluxes obtained for the single stars requested by the theoretical isochrones arise from transformations based on empirical photometric libraries and relations and not from the use of theoretical stellar atmospheres. This is one key difference compared to other models which make use of theoretical stellar atmospheres (see Section \ref{comparison_other_models}).  

\subsection{Quality and range of applicability}\label{Quality}

In this section, we present a way to quantify the quality and reliability of our SSP models as a function of the parameters age and metallicity. To do this, we adopt the normalised quality parameter $Q_n$ introduced in \citet{Vazdekis10} for the model predictions in the optical spectral range. The quality of an SSP depends on the parameter coverage of the input stellar library to be used by the interpolator (see Section \ref{peculiar_stars}) in order to reproduce the single stellar spectra which are later integrated along the isochrone. A large number of stars with parameters differing as little as possible from the ones requested for the computation of one respective model assures a high quality of the resulting model spectrum. The resulting quality value $Q$ for one SSP model is defined as follows:

\footnotesize
\begin{equation}
\resizebox{.9\hsize}{!}
{$ Q=\frac{\sum_{i=1}^{n_{m_t}}\left[\frac{x_s\sum_{j=1}^{8}N_i^j}{\sum_{j=1}^{8}\sqrt{\left(\frac{\Phi_{\theta_i^j}}{\sigma_{\theta_m}}\right)^2+\left(\frac{\Phi_{\log(g)_i^j}}{\sigma_{\log(g)_m}}\right)^2+\left(\frac{\Phi_{\text{[Fe/H]}_i^j}}{\sigma_{\text{[Fe/H]}_m}}\right)^2}}\right]N_iF_{K_i}}{\sum_{i=1}^{n_{m_t}}N_iF_{K_i}}$}.
\label{Quality_parameter_eq}
\end{equation}
\normalsize


In Eq. \ref{Quality_parameter_eq}, $N_i$ designates the total number of stars in the different mass bins ranging from the smallest mass $m_1$ to the most massive stars of mass $m_t$ which can be encountered at the specific age $t$ of the stellar population. In contrast, $N_i^j$ stands for the number of stars which are found in each of the eight cubes in parameter space (see Section \ref{peculiar_stars}). The sizes of these cubes are defined as $\Phi_{\theta_i^j}$, $\Phi_{\log(g)_i^j}$ and  
$\Phi_{\text{[Fe/H]}_i^j}$, respectively, whereas $\sigma_{\theta_m}$,  $\sigma_{\log(g)_m}$ and $\sigma_{\text{[Fe/H]}_m}$ represent the minimum 
uncertainties of these three stellar atmospheric parameters. The minimum cube sizes are obtained by multiplying these $\sigma$ values by $x_s$. $F_{K_i}$ gives the flux in the $K$ band. 
$Q$ is then normalised by dividing it by a minimum value $Q_m$ corresponding to the poorest still acceptable value of $Q$. For $Q_m$, we assume that stars are located in at least three out of the eight cubes and that the maximum cube sizes do not exceed one-tenth of the total parameter ranges of our library. Consequently, models of values of $Q_n$ larger than 1 are assumed to be of sufficient quality to be used safely. The interested reader can find a more detailed description of the definition of this parameter estimating the quality of our SSP models in \citet{Vazdekis10}.\looseness-2

\begin{figure}
\begin{center}
 \resizebox{\hsize}{!}{\includegraphics{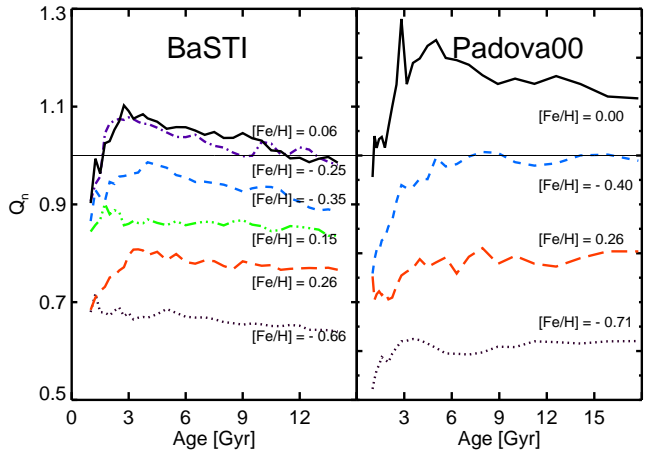}}
  \caption{Normalised quality parameter $Q_n$ as a function of age for SSP models of different metallicities. The horizontal line at $Q_n=1$ marks the general limit of reliability of the models (see text). The left hand panel is dedicated to models based on BaSTI, whereas the right hand panel depicts the quality behaviour of Padova00 based SSP models.}\qquad
\label{Quality_parameter}
\end{center}
\end{figure}

Figure \ref{Quality_parameter} displays the behaviour of the normalised quality parameter $Q_n$ as a function of age for SSP models of different metallicities. In the case of underlying BaSTI isochrones (left panel of Figure \ref{Quality_parameter}), the models of Solar and slightly subsolar metallicities show the highest values of $Q_n$ exceeding the limit of 1 at least for ages larger than $\approx$ 2 Gyr. For Padova00 based models (right-hand panel), only the SSP models of Solar metallicities provide values of $Q_n > 1$ for the whole age range. The quality parameters of the models of ${\rm[Fe/H]=-0.40}$ already lie slightly below 1. Nevertheless, they can be considered acceptable as well. For lower metallicities and also for supersolar values the models are of lower quality and therefore must be handled with some care. This is equally true for the youngest ages between 1 and 2 Gyr. The low quality of the models in this age range can be also understood by the mismatch between the bolometric flux assigned to AGB and carbon stars in this age range and the actually available number of these stars in our input sample. Table \ref{Parameter_coverage} summarises the ranges of the various parameters for which we were able to compute reliable SSP models. Note that the values obtained here for the quality parameter $Q_n$ are much lower than the ones calculated for the model predictions based on MILES \citep[see Figure 6 in][]{Vazdekis10}.


\begin{table*}
\caption{ Properties of the model SSP SEDs}
\label{Parameter_coverage}
\centering
\begin{tabular}{l c}

\hline
Spectral coverage\\
\hline
  Spectral range & ${\rm 0.815-5 \, \mu m}$ \\ 
  Spectral resolution & ${\rm\sigma=60 \,km \, s^{-1}}$, (FWHM varies between ${\rm5.7 \,\AA}$ at ${\rm \lambda=0.945 \, \mu m}$ and ${\rm 16.1 \, \AA}$ at ${\rm \lambda=3.475 \,\mu m}$) \\
  Pixel size  & ${\rm 2.73 \, \AA \cdot pixel^{-1}}$ \\
  \hline 

Parameter coverage\\
\hline
  IMF type & unimodal, bimodal, Kroupa, Kroupa revised\\ 
  IMF-slope (for unimodal and bimodal) & 0.3-3.3 \\
  Isochrones  &  BaSTI \citep{Pietrinferni04}, Padova00 \citep{Girardi00}\\
  metallicity (BaSTI) & ${\rm -0.66 \, dex < [Fe/H] < 0.26 \,dex}$ \\
  metallicity (Padova00) & ${\rm -0.71 \, dex < [Fe/H] < 0.22 \,dex}$ \\
  ages (BaSTI)         & ${\rm 1 \, Gyr < \textit{t} < 14 \,Gyr}$ \\
  ages (Padova00)      & ${\rm 1 \, Gyr < \textit{t} < 17.78 \,Gyr}$\\
\hline
\end{tabular}
\end{table*}

\section{Colours measured on the SSP SEDs}\label{colors}

In this section, we discuss the behaviour of colours measured from our SSP-SEDs as a function of age and metallicity. In order to obtain colours from our SSP model spectra, we convolve the spectral flux with the respective filter response functions. In this procedure, we calibrate the resulting magnitudes based on the Vega-spectrum of \citet{Castelli94} as a zeropoint. We calculated the magnitudes in the 2MASS $J$, $H$, and $K_s$ filters, in the Johnson $J$ and $K$ and in the \textit{Spitzer} [3.6] and ${\rm[4.5] \,\mu m}$ filters. As described in Section \ref{Preparing the stellar library}, we prepared all of the stellar spectra used in the modelling process to end at $5 \, \mu m$. However, the \textit{Spitzer} ${\rm[4.5] \,\mu m}$ band extends slightly beyond ${\rm 5 \, \mu m}$, up to ${\rm 5.22 \,\mu m}$. Consequently, we need to quantify the missing flux in the ${\rm [4.5] \, \mu m}$ band. Therefore, we made use of a number of IRTF stars whose spectra extend well beyond ${\rm5 \, \mu m}$ and calculated the flux in the complete \textit{Spitzer} ${\rm [4.5] \, \mu m}$ filter as well as in a shorter ${\rm[4.5] \, \mu m}$ filter ending at ${\rm 5 \,\mu m}$. The resulting difference in flux was only about ${\rm \Delta([3.6]-[4.5])=0.0001}$ and thus negligible. So, the \textit{Spitzer} (${\rm[3.6]-[4.5]}$) colour determined from our SSP models can be safely compared to the same colour from observations. 

\subsection{Internal consistency of the model colors}

\begin{figure}
\begin{center}
 \resizebox{\hsize}{!}{\includegraphics{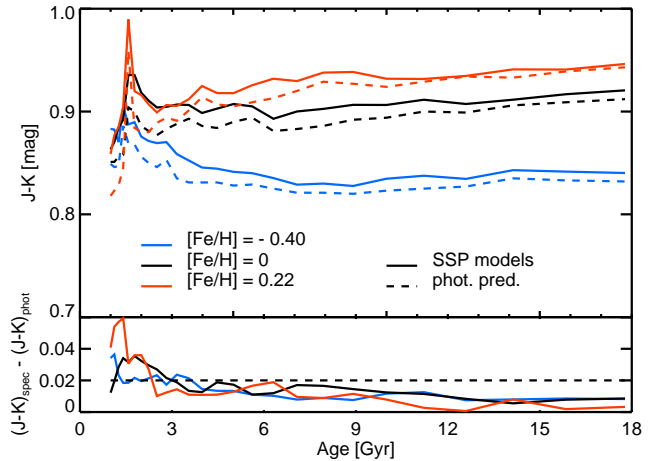}}
  \caption{Comparison between the behaviour with age of the ($J-K$) colour extracted from our SSP models and the same colour predicted from the models based on the MILES stellar library \citep{Vazdekis10}. The dashed lines delineate the MILES photometric predictions and the solid ones the colours calculated from our SSP models. The different colours represent different metallicites (see legend). In the lower panel, we plot the absolute differences between the colours. The dashed line marks the maximum difference of $\Delta = 0.02 \, mag$ which we consider still acceptable.}\qquad
\label{MILES_photometric_predictions}
\end{center}
\end{figure}

In Figure \ref{MILES_photometric_predictions}, we compare the ($J-K$) colour as calculated from our model SSPs with the same colour originating from the so-called MILES photometric predictions. These photometric colour predictions in the NIR have been determined by integrating the stellar fluxes obtained from the metallicity-dependent colour-temperature relations for dwarfs and giants described in Sections \ref{SAPs_color_temperature_relations} and \ref{model_computation}, which are based on extensive empirical photometric stellar libraries. Note that consequently, these colours were not obtained by convolving any synthesized SSP model SEDs with the respective filter responses, since the MILES models extend only up to ${\rm9500 \,\AA}$ \citep{Vazdekis12}. Instead, these colours were calculated by using the metallicity-dependent colour-temperature relations mentioned above.


As can be seen from Figure \ref{MILES_photometric_predictions}, the ($J-K$) colour from the MILES photometric predictions agrees very well with the same colour measured from our SSP models over the whole range of ages, from 1 to 17.78 Gyr. Since these photometric predictions have only been published based on Padova00 (see http://miles.iac.es), we present only comparisons to models using this set of isochrones. As is explicitly shown in the lower panel of Figure \ref{MILES_photometric_predictions}, the colours from our photometric predictions are less than $ {\rm0.02 \,mag}$ bluer than the colours integrated from our SSP spectra for all ages larger than ${\rm 3 \,Gyr}$. This holds for all metallicities within the covered range. For younger ages between 1 and ${\rm 3 \,Gyr}$, the colour differences are slightly larger. This close agreement shows that our models are internally consistent.


\subsection{Sensitivity of the \textit{Spitzer}-$[3.6]-[4.5]$ colour to age and metallicity}\label{age_FeH_behaviour}

\begin{figure}
\begin{center}
 \resizebox{\hsize}{!}{\includegraphics{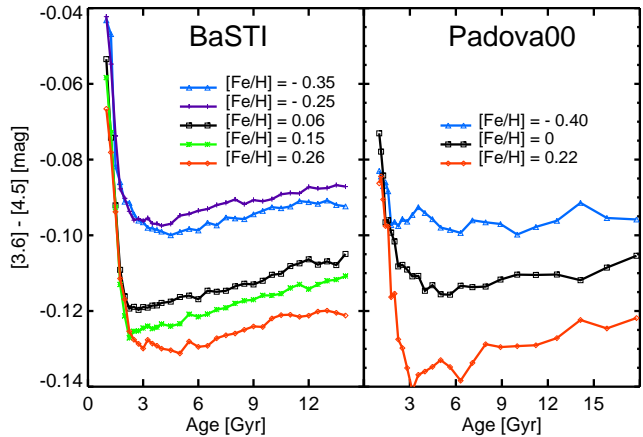}}
  \caption{Behaviour of the \textit{Spitzer} ($[3.6]-[4.5]$) colour determined from our SSP models with age. The different colours and symbols represent different metallicites (see legend). The colour predictions corresponding to the highest and lowest metallicity values are of slightly lower quality. The left panel displays colours from models based on BaSTI, whereas the colours in the right one originate from models based on Padova00.}\qquad
\label{Spitzer_color_versus_age}
\end{center}
\end{figure}

In this Section, we study the behaviour of the \textit{Spitzer} ($[3.6]-[4.5]$) colour as a function of age and metallicity. From Figure \ref{Spitzer_color_versus_age} we see that this colour does not change much for ages above 2 Gyr. For a fixed metallicity, the \textit{Spitzer} $[3.6]-[4.5]$ colour becomes redder with increasing age by around ${\rm0.01 \, mag}$ which is still comparable to the span of colour variation of around ${\rm 0.002 \,mag}$ cited by \citet{Meidt14} over the whole age range. This is equally true for the SSP models based on BaSTI as well as for the ones based on Padova00. For ages smaller than 2 Gyr, we observe a steep rise to redder colours. This is equally true for both the BaSTI and the Padova00 based models, however it is more pronounced reaching redder colours in the former case. This increase in the \textit{Spitzer} $([3.6]-[4.5])$ colour can be explained by the enhanced contribution of AGB stars in this age range (see also fractions in Section \ref{Cool_stars}). The observed difference between the behaviour of the colours originating from the BaSTI and from the Padova00 based models shows how sensitive this age regime is to the modelling of the AGB stars. In addition, the quality parameter (see Section \ref{Quality}) decreases to values smaller than 1 for SSP models of ages younger than 2 Gyr. 

Figure \ref{Spitzer_color_versus_age} also shows that the metallicity dependence of the \textit{Spitzer} ($[3.6]-[4.5]$) colour is rather weak, however clearly visible. Over the whole age range, the overall difference between the colours based on the isochrones of different metallicities equals up to 0.04. Contrary to the behaviour of the ($J-K$) colour with metallicity (see Figure \ref{MILES_photometric_predictions}), the more metal-poor SSP models give slightly redder \textit{Spitzer} ($[3.6]-[4.5]$) colours compared to the more metal-rich ones. A likely explanation for this reversed trend is the presence of the CO absorption band in the ${\rm[4.5]\, \mu m}$ \textit{Spitzer} filter (see also Section \ref{Preparing the stellar library}). This observed trend is also seen by \citet{Peletier12, Meidt14} and \citet{Norris14}, who relate it to the absorption by the CO in the ${\rm[4.5] \, \mu m}$ band. Larger metallicities lead to increased CO absorption, and subsequently to a smaller flux in the ${\rm[4.5] \,\mu m}$ band. This, however, makes the \textit{Spitzer} ($[3.6]-[4.5]$) colour slightly bluer compared to the same colour measured from SSP models of lower metallicities and thus with less CO absorption. This trend just does not hold for the models of ${\rm [Fe/H] = - 0.35}$ from which we measure a \textit{Spitzer} ($[3.6]-[4.5]$) colour which is slightly bluer than the one determined from the models of ${\rm [Fe/H] = - 0.25}$. The difference however is small (${\rm <= 0.01 \, mag}$). Since also the quality of the models of a metallicity of ${\rm [Fe/H] < -0.3 \, mag}$  is only modest (see Figure \ref{Quality_parameter} in Section \ref{Quality}), we are not in a position to decide whether we observe here a real saturation of the \textit{Spitzer} ($[3.6]-[4.5]$) colour with decreasing metallicity or not. Our data do not permit us to investigate in detail the depth of the CO band as a function of metallicity, as the IRTF spectra show a gap due to telluric absorption in this range, which was corrected artificially using the theoretical Phoenix library (see Section \ref{Preparing the stellar library}). Yet a comparison to the colours obtained from other (theoretical) models in this wavelength range which reach lower metallicities seems to indicate a continuation of this trend of reddening \citep[][, see below]{Norris14}. 



The small metallicity dependence of the \textit{Spitzer} ($[3.6]-[4.5]$) colour is in agreement with the literature. \citet{Meidt14} give a variation of $\Delta([3.6]-[4.5])=0.06$ over the larger metallicity range from ${\rm[Fe/H] = -1.7}$ to ${\rm[Fe/H] = 0.4}$. This corresponds to $\Delta([3.6]-[4.5]) = 0.01$ over the metallicity range ${\rm \Delta[Fe/H] = 0.35}$, which is the range of highest reliability of our models. \citet{Peletier12} state that barely any fluctuation of the \textit{Spitzer} ($[3.6]-[4.5]$) colour is visible for metallicities higher than $Z=0.008$, which equals ${\rm[Fe/H] = -0.4}$.

\begin{figure}
\begin{center}
 \resizebox{\hsize}{!}{\includegraphics{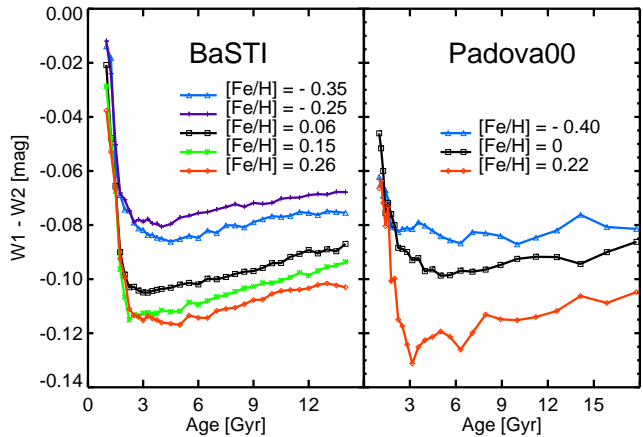}}
\caption{Same as Figure \ref{Spitzer_color_versus_age}, but for the \textit{WISE} ($W1-W2$) colour.}\qquad
\label{WISE_color_versus_age}
\end{center}
\end{figure}

Finally, we integrate the spectral flux from our models falling within the two filters W1 and W2 of the Wide-field Infrared Survey Explorer (WISE) satellite \citep{Wright10} and thus obtained the \textit{WISE} $(W1-W2)$ colour. The \textit{WISE} W1 and W2 filters are only slightly shifted with respect to the \textit{Spitzer} ${\rm[3.6]\,\mu m}$ and ${\rm[4.5]\,\mu m}$ bands. Subsequently, studying the behaviour of the \textit{WISE} ($W1-W2$) colour with age (see Figure \ref{WISE_color_versus_age}) allows us to confirm the trends observed for the \textit{Spitzer} colour (see Figure \ref{Spitzer_color_versus_age}). Indeed, when comparing Figure \ref{Spitzer_color_versus_age} to Figure \ref{WISE_color_versus_age}, we can conclude that these two colours do behave basically the same as a function of age and metallicity. Due to the slightly different filter responses which are shifted with respect to each other, the \textit{WISE} ($W1-W2$) colour is generally redder by about ${\rm 0.03 \, mag}$ for ages younger than 2 Gyr and by still about ${\rm 0.01 - 0.02 \, mag}$ for older ages. Furthermore, the \textit{WISE} W2 filter seems to be slightly less affected by the CO absorption band centred at ${\rm 4.6 \,\mu m}$, since the blueing of colours with increasing metallicity is a bit less pronounced. In a recent publication, \citet{Norris14} have determined \textit{WISE} photometry of a diverse sample of globular clusters and early-type galaxies and then compared the resulting $(W1-W2)$ colour to the predictions of some stellar population models (see next Section \ref{comparison_other_models}). As can be seen from their Figure 4, their observed sample shows the same blueing of the \textit{WISE} $(W1-W2)$ colour with increasing metallicity which we observe from our models. Apparently, this trend holds even for lower metallicities which we cannot reach with our models due to the limitation of the IRTF library.

\section{Comparing to other SSP models from the literature}\label{comparison_other_models}

In this Section we compare our predictions with other models available in the literature. According to \citet{Barmby12}, there exist only three other models which provide explicit colour predictions in the wavelength range beyond the $K$ band. These three are the SSP models of the Padova group emerging as a by-product of their library of stellar isochrones \citep[][hereafter Marigo-models]{Marigo08, Girardi10}, the GALEV models \citep{Kotulla09} and the FSPS ones \citep{Conroy09, Conroy10}. However, it should be mentioned that IR colours can be also obtained from all other models which are based on the theoretical BaSeL library \citep{Kurucz92, Lejeune97, Lejeune98, Westera02} like e.g. \citet{Bruzual03} and \citet{Maraston05} by convolving the predicted spectra with the respective filter response functions. In what follows we focus however on the three models giving directly colour predictions in the IR discussed also by \citet{Barmby12} complemented by the predictions of the newest release of the Padova models by \citet{Bressan12}.

All of these models are based on Padova isochrones. However, the models differ from each other in the stellar libraries on which they are based. Contrary to our models, all of the underlying stellar libraries are theoretical ones. So, the GALEV models use version 2.2 of the theoretical BaSeL stellar library \citep{Lejeune98}, whereas the FSPS models rely on the updated version 3.1 of this same spectral library \citep{Westera02} in the wavelength range between ${\rm 2.5 \, \mu m}$ and ${\rm 5 \, \mu m}$. The spectra of the BaSeL library are based on stellar atmospheres from \citet{Kurucz92} and - in the case of M giants - from \citet{Fluks94} and \citet{Bessell89, Bessell91}. 
The Marigo models mainly make use of the revised ATLAS9 model spectra from \citet{Castelli03}. For M, L and T dwarfs, they adopt the model atmospheres of \citet{Allard00} and for M giants those of \citet{Fluks94}.
Moreover, the prescriptions for the thermally pulsating asymptotic giant branch (TP-AGB) stars, vary between the models. In order to account for them, FSPS uses empirical spectra from \citet{Lancon02} and the Marigo models synthetic spectra from \citet{Loidl01}. As these two spectral libraries extend only up to ${\rm [2.5] \, \mu m}$, they have to be extrapolated with a Rayleigh-Jeans model. For carbon stars, the FSPS models are based on the models by \citet{Aringer09}. The GALEV models do not treat post-main sequence stars in a particular way.
For all of these four families of models, we obtained the \textit{Spitzer} ${\rm [3.6] \,\mu m}$ and ${\rm [4.5] \, \mu m}$ magnitudes for a Solar metallicity and for all available ages older than 1 Gyr. The models adopted by us are based on a Kroupa-like IMF. 


In the case of the GALEV models, we used the default parameters and selected to neglect the dust contribution. For the Marigo models, we used the evolutionary tracks by \citet{Marigo08} together with the \citet{Girardi10} case A correction for low-mass, low-metallicity AGB tracks, whereas we used version 1.1 of the PARSEC models of \citet{Bressan12}. For both sets of models, we did not select any circumstellar dust. In the case of FSPS, we retrieved the models made readily available online based on Padova isochrones and the BaSeL stellar library. 

\begin{figure}
\begin{center}
\resizebox{\hsize}{!}{\includegraphics{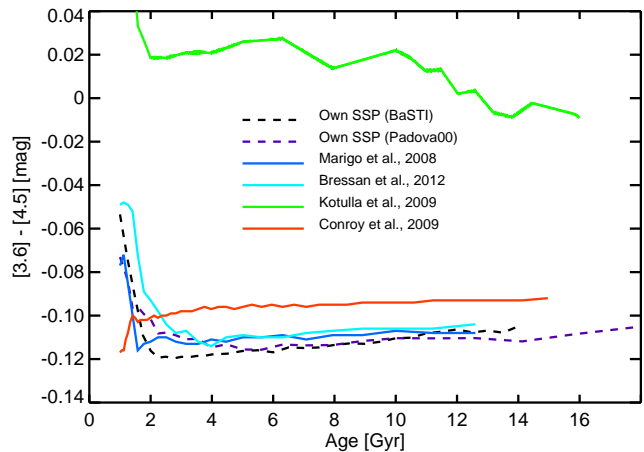}}\qquad
\caption{Comparison of the behaviour of the \textit{Spitzer} ($[3.6]-[4.5]$) colour with age for four different models (see legend) and our own (BaSTI- and Padova00 based) ones. All of the displayed models are of Solar metallicity.}\qquad
\label{Comparison_other_models}
\end{center}
\end{figure}

Figure \ref{Comparison_other_models} shows the behaviour of all these models together with our new SSP models based on BaSTI and Padova00 isochrones, respectively, for ages older than 1 Gyr. The \textit{Spitzer} ($[3.6]-[4.5]$) colours predicted by the Marigo models and the Bressan models agree almost perfectly with our two new sets of models over the whole age range. For ages older than about 2 Gyr, the Marigo and the Bressan models give colors which are generally only around ${\rm 0.005 \, mag}$ redder than those of our two sets of models. 

The \textit{Spitzer} ($[3.6]-[4.5]$) colours derived from the FSPS models also coincide reasonably well with the colours determined from our sets of SSP models and the Marigo and Bressan ones. The almost constant colours obtained from FSPS are about ${\rm 0.01 - 0.015 \,mag}$ redder than the ones measured from the former ones for ages older than 2 Gyr. For the youngest ages, the FSPS models do not show the turn to redder colours due to the enhanced contribution of AGB stars which becomes visible in our models and in the Padova ones \citep[compare also][]{Maraston98, Maraston05}.

The GALEV models, however, give absolute values of the \textit{Spitzer} ($[3.6]-[4.5]$) colour which are completely offset to redder colors by about $0.10-0.13$ from our SSP models. Most likely, there are two main reasons why this set of models fails to provide colour predictions which lie in the same range as all the others, although their input parameters do not differ much from e.g. the Marigo or Bressan models \citep[see above and][]{Barmby12}. On the one hand, their far-too-red colours might be explained by their choice of theoretical stellar atmospheres for transforming the theoretical stellar parameters to the observational plane. On the other hand, they do not take the CO absorption in the \textit{Spitzer} ${\rm[4.5] \,\mu \, m}$ band into account. The crucial point of correcting for this absorption in order to obtain reliable colour predictions in the IR is also pointed out by \citet{Norris14} who compare their observations also to the GALEV models (see their Figure 4). They claim that the FSPS models of \citet{Conroy09, Conroy10} do not consider the CO absorption neither, a finding that we cannot confirm from the predictions visualised in our Figure \ref{Comparison_other_models}. Apart from that, \citet{Norris14} state that the models by \citet{Bressan12} are the first modern SSP models which are able to reproduce the observed $(W1 - W2)$ colours of dust-free stellar populations. Since the colour predictions arising from our SSP models closely agree with the ones of \citet{Bressan12}, we can argue that this also holds for our newly calculated models based on empirical stellar spectra.

 

\section{Mass-to-light ratios in the Spitzer [3.6] and [4.5] bands and dependence on the IMF}\label{M_L_IMF}

In this Section, we study the behaviour of the mass-to-light ratios in the \textit{Spitzer} ${\rm[3.6] \, \mu m}$ and ${\rm[4.5] \, \mu m}$ bands as a function of age and metallicity. We also investigate the impact of varying the IMF on the mass-to-light ratios in those two bands. As expected, here the effect of the IMF is much more pronounced than for the colours. Nevertheless, we also discuss briefly the IMF dependence of the \textit{Spitzer} ($[3.6]-[4.5]$) colour (see Subsection \ref{IMF}).

\subsection{Mass-to-light ratios in the IR}\label{M_L}

Apart from the spectra of the SSPs, our models also give their stellar masses as an output. Together with the flux in a specific band which can be obtained by integrating over the model spectra, we are subsequently able to calculate mass-to-light ratios in various bands. In what follows, we determine the mass-to-light ratios in the Johnson $K$ band as well as in the \textit{Spitzer} ${\rm [3.6] \, \mu m}$ and ${\rm [4.5] \, \mu m}$ bands. We provide all of our mass-to-light ratios in Solar units, $M_{\odot}/L_{\odot}$. Therefore, we adopted the absolute Solar magnitudes of $M_{\odot}^{3.6} = 3.24$ and $M_{\odot}^{4.5} = 3.27$ from \citet{Oh08} and $M_{\odot}^{K} = 3.28$ from \citet{Binney98} in order to calculate the luminosities. 

The mass-to-light ratios measured in these NIR and IR bands present the great advantage that they depend less on parameters like age and metallicity and hence are more stable and constant compared to their counterparts determined in the optical regime. \citet{Bell01}, e.g., mention that over the same parameter range, the mass-to-light ratios change by a factor of about 7 in the $B$ band, 3 in the $I$ band and 2 in the $K$ band. This behaviour of the mass-to-light ratio can be understood considering that the light of SSPs in the IR is dominated by red giant branch (RGB) stars. With advancing time, the RGB gets more and more populated by stars of later spectral type with decreasing mass-to-light ratios. At the same time, however, the mass-to-light ratio of the remaining main sequence stars increases since it is mainly determined by later type M stars of higher mass-to-light ratio. Consequently, the overall changes in the mass-to-light ratio with time are rather smooth as these two effects compensate each other \citep[see, e.g.,][]{Meidt14}. 

\begin{figure}
\begin{center}
 \resizebox{\hsize}{!}{\includegraphics{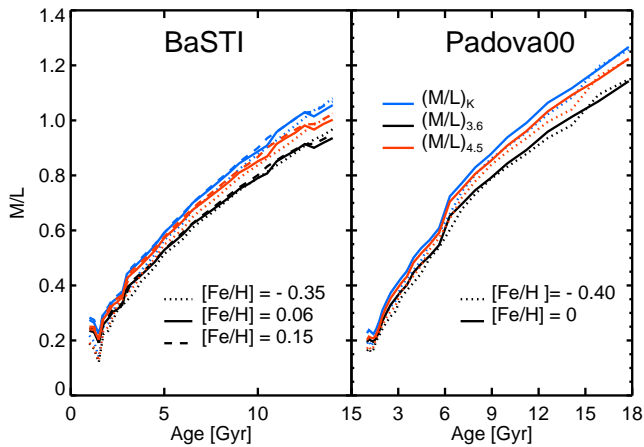}}
  \caption{Behaviour of various $M/L$ ratios with age. Blue lines represent the $M/L$ ratio in the $K$ band, black lines in the ${\rm [3.6] \, \mu m}$ band and red lines at ${\rm [4.5] \, \mu m}$. The different line styles stand for the metallicities as indicated in the legend. The $M/L$ ratios plotted in the left hand panel originate from SSP models based on BaSTI isochrones, whereas the ones shown in the right hand panel were obtained from SSP models based on Padova00 ones.}\qquad
\label{M_L_ratio_compare}
\end{center}
\end{figure}

Such a generally small parameter dependence is also confirmed by Figure \ref{M_L_ratio_compare}, depicting the $M/L$ ratios in the three NIR and IR bands and for three (in the case of SSP models based on BaSTI isochrones) and two (SSP models based on Padova isochrones) different metallicities. In particular, there is basically no dependence on metallicity. This insensitivity to metallicity is also confirmed by \citet{Norris14} who calculate $M/L$ ratios in the \textit{WISE} W1 and W2 bands from the newest version of the Padova models \citep{Bressan12}. The differences between the $M/L$ ratios in models of varying metallicity are smaller than the differences between the $M/L$ ratios measured in the various filters. However, the dependence with age is smallest for the $M/L$ ratio in the ${[3.6] \, \mu m}$ band, whereas the time evolution of the $M/L$ ratio in the ${[4.5] \, \mu m}$ band resembles a lot the one in the $K$ band (see again Figure \ref{M_L_ratio_compare}). The $M/L$ ratios of the two latter ones are slightly larger than the one measured in the ${\rm [3.6] \, \mu m}$ band. In general, the $M/L$ ratio determined in the ${\rm [3.6] \, \mu m}$ \textit{Spitzer} band just changes from $(M/L)_{3.6}\approx0.20$ for an age of 1 Gyr to $(M/L)_{3.6}\approx1.1$ at the maximum age of ${\rm 17.78 \, Gyr}$, i.e., by a factor of 5.5. \citet{Norris14} find an increase of the $M/L$-ratio in the \textit{WISE} bands by a factor of 2 between 3 and 10 Gyr which agrees perfectly with our results (compare Figure \ref{M_L_ratio_compare} to their Figure 5). In contrast to this, the $M/L$ ratio in the $V$ band calculated from MILES SSP models \citep{Vazdekis10} of Solar metallicity grows from $(M/L)_V\approx0.6$ to $(M/L)_V\approx6$ within the same age range, i.e., it increases roughly by a factor of 10.

Furthermore, comparing the left and right hand panels of Figure \ref{M_L_ratio_compare}, it can be seen that the$ M/L$ ratio mildly depends on the employed isochrones.

\begin{figure}
\begin{center}
 \resizebox{\hsize}{!}{\includegraphics{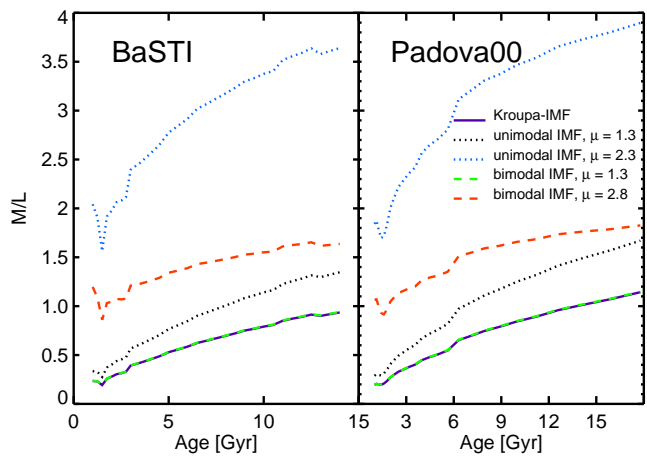}}
  \caption{Behaviour of various $M/L$ ratios at ${\rm[3.6] \, \mu m}$ with age. The depicted $M/L$ ratios originate from SSP models of Solar metallicities calculated using different IMFs (see legend). The $M/L$ ratios plotted in the left-hand panel are calculated from SSP models based on BaSTI isochrones, whereas the ones shown in the right-hand panel were obtained from SSP models based on Padova00 isochrones.}\qquad
\label{M_L_ratio_compare_IMF}
\end{center}
\end{figure}

On the contrary, the $M/L$ ratios are highly sensitive to the IMF used in the modelling procedure, as Figure \ref{M_L_ratio_compare_IMF} clearly depicts. In general, $M/L$ ratios originating from SSP models based on unimodal IMFs are larger than the ones based on bimodal IMFs. This is due to the enhanced contribution of faint low-mass stars in the case of unimodal IMFs compared to bimodal IMFs for which the number of stars with masses smaller than ${\rm0.6 \,M_{\odot}}$ is reduced \citep[e.g., Figure 2 in][]{LaBarbera13}. Also note that a bimodal IMF of slope 1.3 is virtually identical to the Kroupa IMF. Moreover, it can be seen from Figure \ref{M_L_ratio_compare_IMF} that a steepening in IMF slope leads to a greater increase in the $M/L$ ratio for unimodal IMFs than for bimodal ones. According to \citet{LaBarbera13}, the $M/L$ ratios arising from models based on unimodal IMFs give too large values, exceeding by far dynamically determined $M/L$ ratios like, e.g., those of \citet{Cappellari13}. Stellar $M/L$ ratios calculated based on bimodal IMFs, however, are consistent with the latter ones also in the most bottom-heavy cases \citep[i.e., for the largest slopes, see also][]{Spiniello14}.

\citet{Meidt14} calculated $(M/L)_{3.6}$ ratios based on the model predictions of \citet{Bruzual03} which they then transformed to ($[3.6]-[4.5]$) using a newly calibrated relation between the NIR and the IR colors of giants observed with GLIMPSE \citep{Benjamin03, Churchwell09}. They find only very small variations of the $(M/L)_{3.6}$ ratios as a function of age and metallicity. That is why they suggest to use even a constant $M/L$ ratio of $M/L_{3.6}\approx0.6$ independent of the \textit{Spitzer} ($[3.6[-[4.5]$) colour and consequently of age and metallicity \citep{Meidt14}. Recently, \citet{Norris14} confirmed that based on this constant $M/L$ ratio, it is possible to determine stellar masses with errors no larger than ${\rm 0.1 \, dex}$ in the middle of the age range 3 to 10 Gyr. The value of $M/L_{3.6}\approx0.6$ is also quite well compatible with an average value for our $M/L$ ratios. However, we obtain a larger dependency on the two parameters age and metallicity, which is also due to the fact that our models are single-burst like and do not contain exponentially declining star formation histories \citep{Norris14}.

\subsection{IMF dependence of our models}\label{IMF}

Up to here, we have always discussed the SSP models based on a Kroupa-like IMF. However, as we pointed out in Section \ref{model_ingredients}, we also calculated our SSP models for a suite of IMF shapes and slopes. The two sets of IMFs which we used were the unimodal and the bimodal ones \citep[see][]{Vazdekis96}. The varying behaviour of a unimodal and bimodal IMF of the same slope $\mu$ for stellar masses smaller than around ${\rm 0.5 \, M_{\odot}}$ is visualized in Figure 3 of \citet{LaBarbera13}. 

\begin{figure*}
\begin{center}
\resizebox{\hsize}{!}{\includegraphics{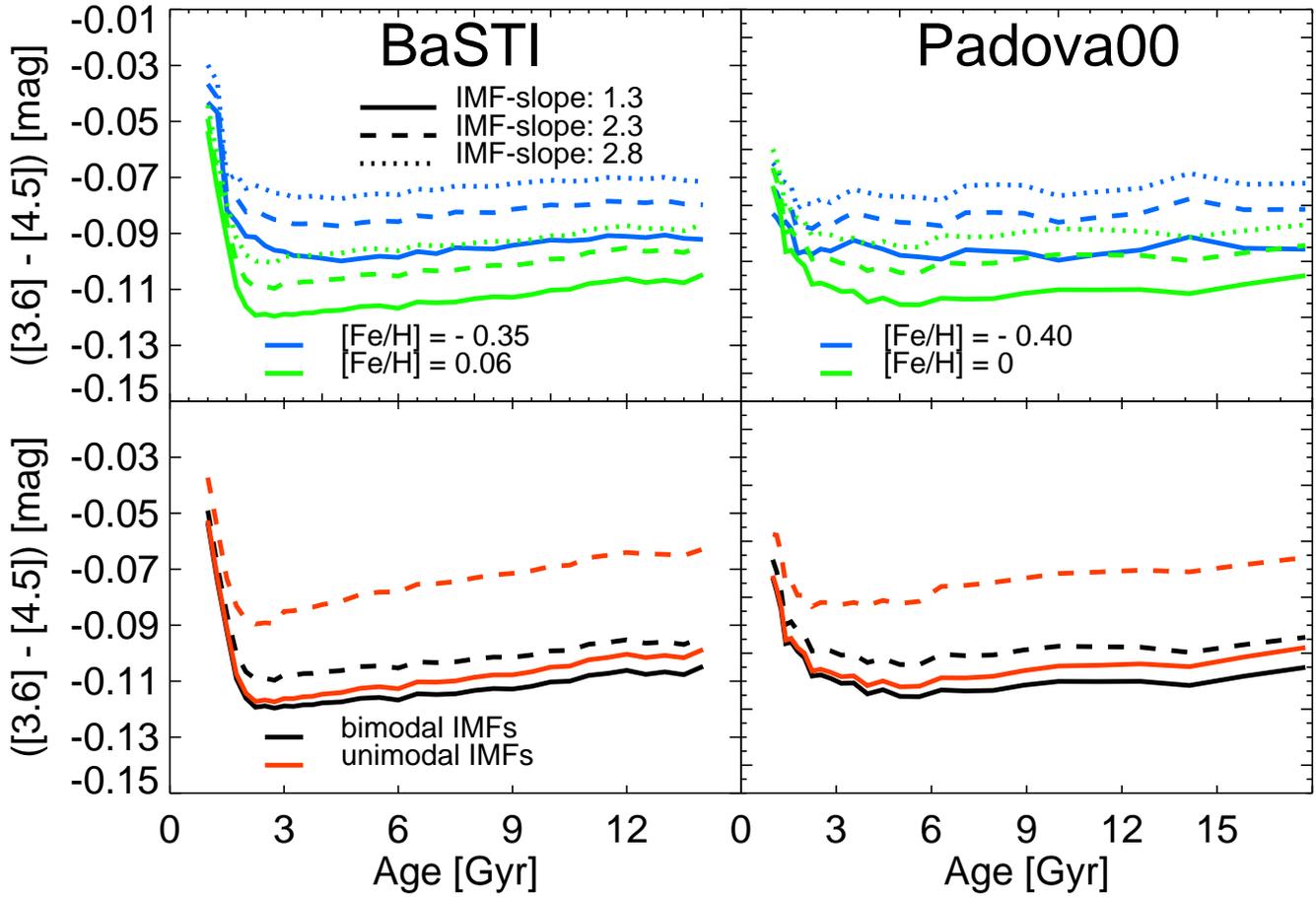}}
  \caption{Behaviour of the \textit{Spitzer} ($[3.6]-[4.5]$) colour as a function of age for SSP models of different IMF types and slopes. The upper left and right panels display the \textit{Spitzer} ($[3.6]-[4.5]$) colour for bimodal IMFs of the three different slopes $\mu=1.3, 2.3, 2.8$ (see also legend). The blue lines show the predicted colours for a subsolar metallicity and the green lines for the Solar one, respectively. Models based on BaSTI are presented in the left-hand panel, whereas Padova00 based models are shown in the right-hand one. The lower two panels visualise the \textit{Spitzer} ($[3.6]-[4.5]$) colour originating from SSP models based on Solar metallicities and uni- and bimodal IMFs of the two slopes $\mu=1.3, 2.3$.}\qquad
\label{Spitzer_age_varying_IMF}
\end{center}
\end{figure*}

The two upper panels of Figure \ref{Spitzer_age_varying_IMF} show the effect of varying the IMF slope in the case of bimodal IMFs. The corresponding figure for unimodal IMFs is not shown, but looks essentially the same. The general trend that becomes visible is a slight reddening of the \textit{Spitzer} ($[3.6]-[4.5]$) colour with increasing IMF slope of the order of $ 0.02 - 0.03$. However, the difference in colour between two models differing by ${\rm \Delta([Fe/H])\approx 0.40}$ is comparable to the difference between two models of the same metallicity but a $\Delta(\mu)$ of $1-1.5$. Consequently, we can conclude that the effect of varying the IMF slope $\mu$ on the colours is small for our models. 

In the two lower panels of Figure \ref{Spitzer_age_varying_IMF} we compare the colour predictions from SSP models based on a bimodal and a unimodal IMF, respectively, for the same slopes. We find that a bimodal and a unimodal IMF of slope 1.3 give virtually the same \textit{Spitzer} ($[3.6]-[4.5]$) colour over the whole age range. In the case of the steeper slope of 2.3, a small difference of again $ 0.02 - 0.03$ arises. The colours deduced from SSP models based on unimodal IMFs become increasingly redder compared to the ones calculated from models with underlying bimodal IMFs. However, the observed effect is again rather small. 

Finally, we can summarise that the impact of varying both the IMF type and slope on the behaviour of the \textit{Spitzer} ($[3.6]-[4.5]$) colour measured from our SSP models is rather small. The observed dependences described above are of the same order as those seen when varying age and metallicity for ages older than ${\rm \sim 2 \,Gyr}$ and over our limited range of metallicity (see Section \ref{colors}). In general, we can conclude that the \textit{Spitzer} ($[3.6]-[4.5]$) colour can be used to a certain extent as a proxy for distinguishing populations of ages younger than ${\rm 2\, Gyr}$ from older ones and also more metal-rich populations from more metal-poor ones. However, it remains very difficult to make a clear statement about the underlying IMF of a stellar population based on this colour. 

\section{Comparison to observations}\label{Comparison_observations}

In this Section, we carry out a comparison between the colours determined from our SSP models and observed photometry to validate our models. We try to reproduce the colours (see Section \ref{early-type_galaxies}) and the colour-velocity dispersion relations (see Section \ref{colour-velocity_dispersion_relations}) of early-type galaxies. For this, we make use of the sample of 48 galaxies observed with the Spectrographic Areal Unit for Research on Optical Nebulae (SAURON) \citep{Bacon01, deZeeuw02}.

\subsection{Early-type galaxies}\label{early-type_galaxies}

\begin{figure}
\begin{center}
\resizebox{\hsize}{!}{\includegraphics{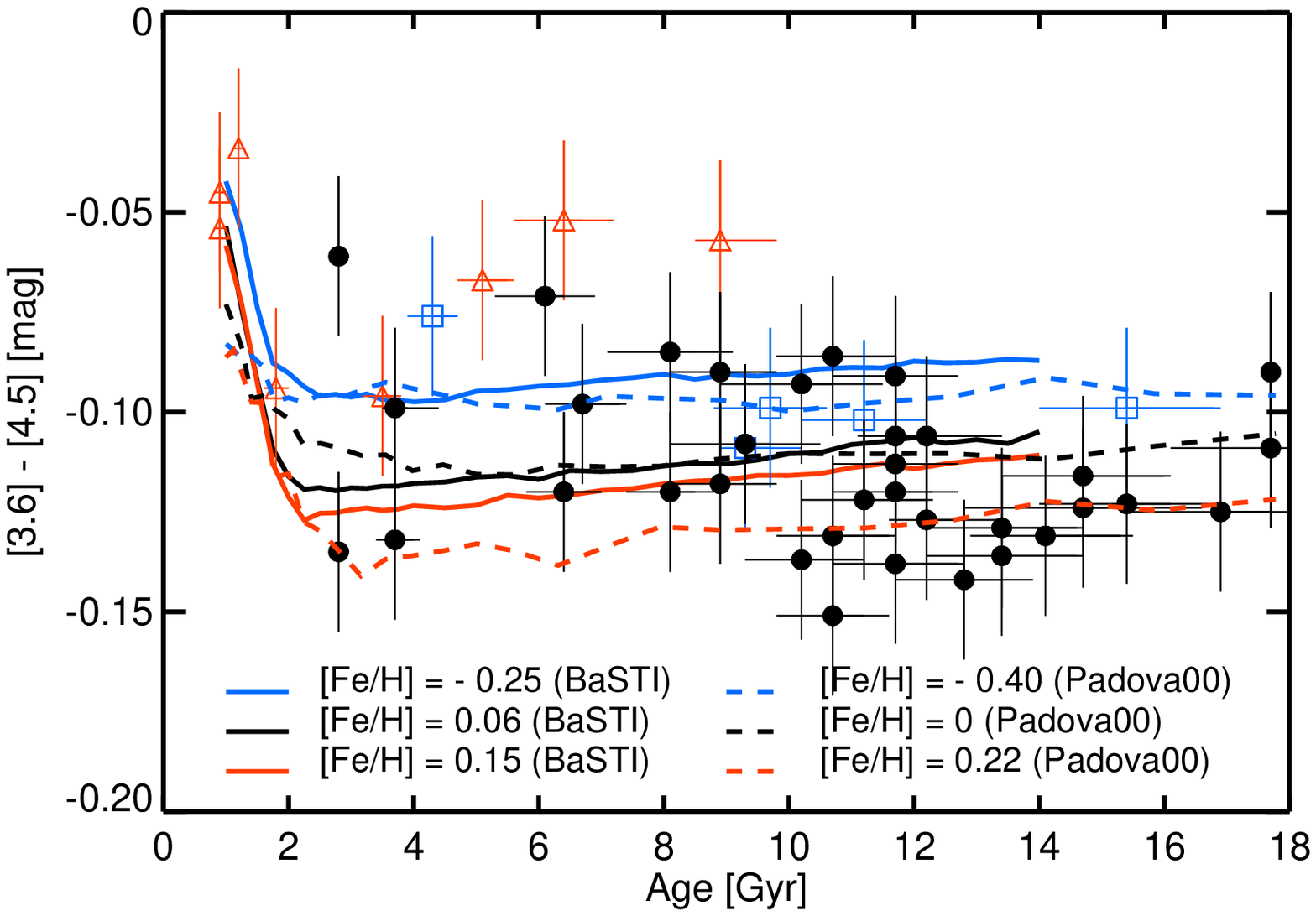}}
\caption{\textit{Spitzer} ($[3.6]-[4.5]$) colour versus age for 48 nearby early-type elliptical and lenticular galaxies from the SAURON survey \citep{Kuntschner10}. Errorbars denote the uncertainties of both colour and age. Galaxies with clear and possible signs of ongoing star formation are shown as red triangles and green squares, respectively, whereas those without star formation are delineated as black dots. The coloured lines show the predictions from our SSP models for various metallicities. Solid lines correspond to models based on BaSTI, whereas the dashed ones delineate Padova00 based models.}\qquad
\label{SAURON_Spitzer_age}
\end{center}
\end{figure}

\begin{figure}
\centering
\resizebox{\hsize}{!}{\includegraphics{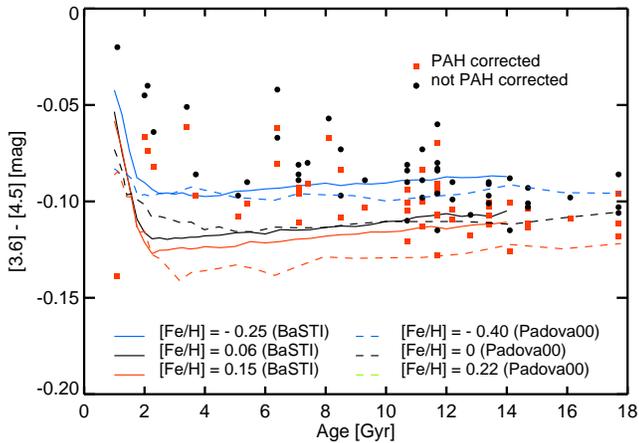}}
\caption{Same as Figure \ref{SAURON_Spitzer_age}, but here age and \textit{Spitzer} colour are determined inside one effective radius. The black dots delineate the SAURON galaxies before they were corrected for PAH emission. The PAH-corrected galaxies are displayed with red squares.}\qquad
\label{SAURON_Spitzer_age_PAH}
\end{figure}

Figure \ref{SAURON_Spitzer_age} compares the observed \textit{Spitzer} ($[3.6]-[4.5]$) colour of the 48 nearby, early-type elliptical and lenticular galaxies from the SAURON survey \citep{Bacon01, deZeeuw02} to the predicted colours of our SSP models as a function of age. The age here is the optical SSP age determined from the SAURON data within $1/8\cdot r_{\text{eff}}$ by \citet{Kuntschner10}. These data show that galaxies become gradually bluer as they become more massive. For this plot, the \textit{Spitzer} colours were measured inside $1/8\cdot r_{\text{eff}}$, rendering them bluer than when determined inside the whole effective radius $r_{\text{eff}}$. Both effects are consistent with metallicity decreasing outwards in the galaxies, causing the CO band to weaken and the ($[3.6]-[4.5]$) colour to redden. 
Our models match quite well the observed colours of the galaxies with the oldest mean luminosity-weighted ages, as determined from the optical spectral range \citep{Kuntschner10}. However, they are unable to reproduce the redder colours of younger galaxies. These galaxies are likely to have more extended SFHs and thus cannot be modelled as a single stellar population. This is confirmed by the classification of \citet{Shapiro10} who estimate the level of star formation present in the SAURON galaxies based on the non-stellar dust emission in the ${\rm [8] \, \mu m}$ \textit{Spitzer} band. The galaxies for which they report clear and likely star formation, respectively, coincide with the reddest galaxies in Figure \ref{SAURON_Spitzer_age}. 
Most of the non-stellar emission in the \textit{Spitzer} ${\rm [3.6] \, \mu m}$ and ${\rm [4.5] \, \mu m}$ bands is due to polycyclic aromatic hydrocarbons (PAHs). \citet{Schoedel14} studied the ratios of the diffuse emission, i.e., the remaining emission after point source subtraction, in the \textit{Spitzer} ${\rm [3.6] \, \mu m}$, ${\rm [4.5] \, \mu m}$ and ${\rm [8] \, \mu m}$ bands for observations of the Galactic centre. They estimate a PAH contribution of 0.03 times the flux at ${\rm [8] \, \mu m}$ in Jansky for the ${\rm [3.6] \, \mu m}$ band and of 0.04 times the ${\rm [8] \, \mu m}$ flux for the ${\rm [4.5] \, \mu m}$ band. We used these estimates in order to correct the \textit{Spitzer} ($[3.6]-[4.5]$) colours of the SAURON galaxies for PAH emission. As can be seen from Figure \ref{SAURON_Spitzer_age_PAH}, we end up with slightly bluer colours which are in better accordance with our models. This is obviously only a crude estimate. Moreover, the colours displayed in Figure \ref{SAURON_Spitzer_age_PAH} were measured within one effective radius contrary to Figure \ref{SAURON_Spitzer_age}.

\begin{figure}
\centering
\resizebox{\hsize}{!}{\includegraphics{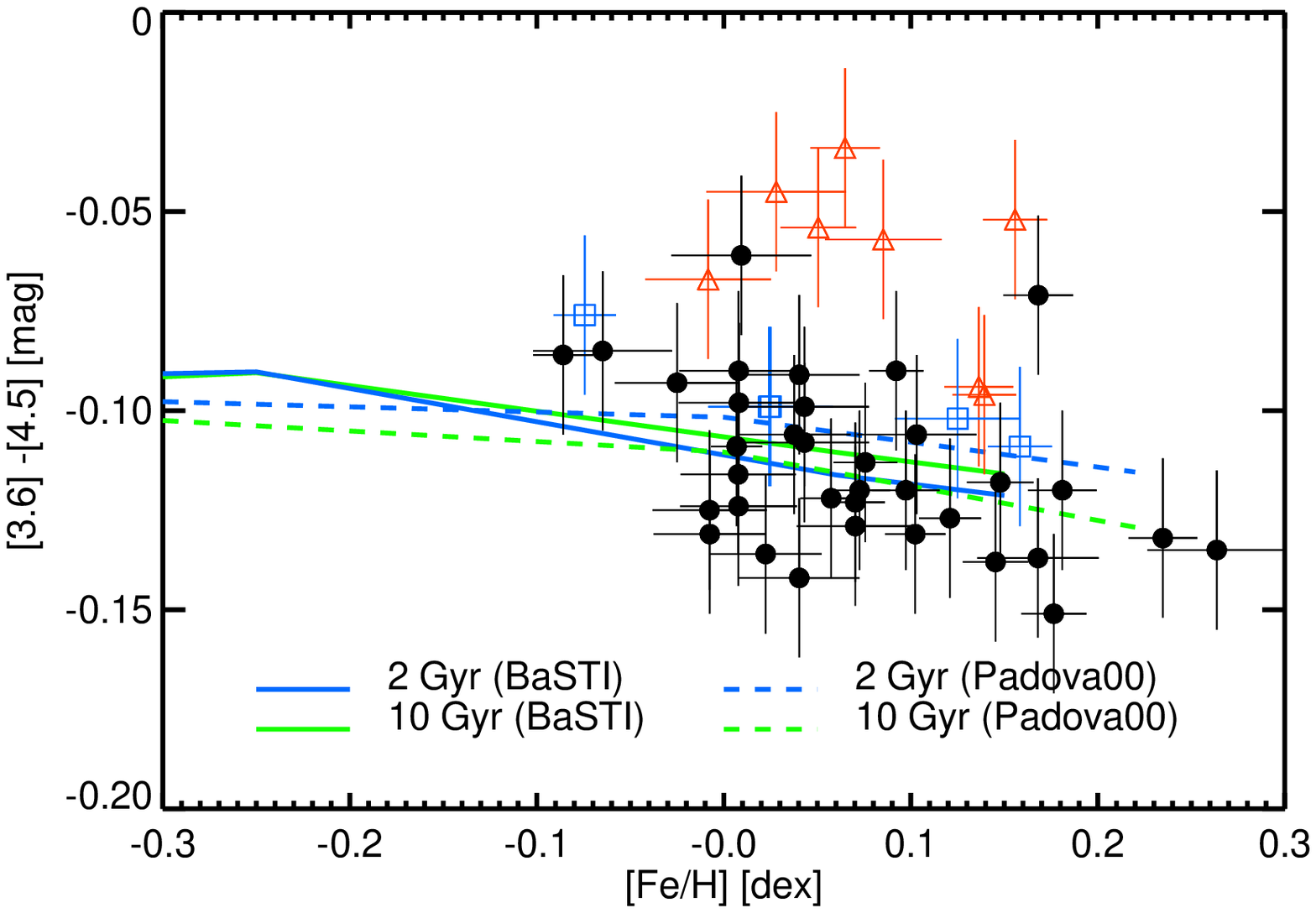}}
\caption{Same as Figure \ref{SAURON_Spitzer_age}, but now plotted versus metallicity. Errorbars denote the uncertainties of both colour and metallicity. The blue and green lines depict SSP models of 2 and ${\rm 10 \,Gyr}$, respectively.}\qquad
\label{SAURON_Spitzer_metallicity}
\end{figure}

Figure \ref{SAURON_Spitzer_metallicity} displays the \textit{Spitzer} colours of the same galaxies, but this time plotted versus metallicity, which was also determined from the SAURON spectra by \citet{Kuntschner10}. Like in the previous Figure \ref{SAURON_Spitzer_age}, the colours and metallicities were measured within $1/8\cdot r_{\text{eff}}$. 
We overplot two models of ages 2 and ${\rm10 \, Gyr}$, based on both BaSTI and Padova00, respectively. Again, our SSP models coincide best with the bluest of the SAURON galaxies, whereas the redder, younger, mostly star-forming galaxies lie outside the range traceable by our models. Figure \ref{SAURON_Spitzer_metallicity} hints that the observed redder ($[3.6]-[4.5]$) colours are rather caused by extended SFHs including also younger populations (compare to Section \ref{age_FeH_behaviour}) than due to lower metallicities. After all, according to Figure \ref{SAURON_Spitzer_metallicity}, the metallicities of the SAURON early-type galaxies are at most slightly subsolar. However, in this paper, we refrain from drawing a final conclusion concerning this point. 

\subsection{Confirming the observed colour-velocity dispersion relations of galaxies}\label{colour-velocity_dispersion_relations}

\citet{Ferreras13} and \citet{LaBarbera13} analyse a variety of optical IMF-sensitive spectral features in a large sample of early-type galaxies and draw conclusions on the underlying IMF. The former ones additionally carry out a spectral fitting using the MIUSCAT population synthesis models \citep{Vazdekis12}. Both find clear evidence for a relation between the IMF and the central velocity dispersion $\sigma_0$. According to their studies, galaxies of ${\rm \sigma_0 \approx 120 \,km \,s^{-1}}$ are characterised by a standard Kroupa-like IMF, whereas galaxies showing a central velocity dispersion of  ${\rm\sigma_0 \approx 200 \,km \,s^{-1}}$ and even larger, are best described by a bottom-heavy bimodal IMF with slope of $\mu > 2$. This behaviour is visualised in Figure 12 of \citet{LaBarbera13} and in Figure 4 of \citet{Ferreras13}. \citet{Spiniello14} find a similar correlation between the central velocity dispersion and the IMF slope for early-type galaxies.

\begin{figure}
\begin{center}
 \resizebox{\hsize}{!}{\includegraphics{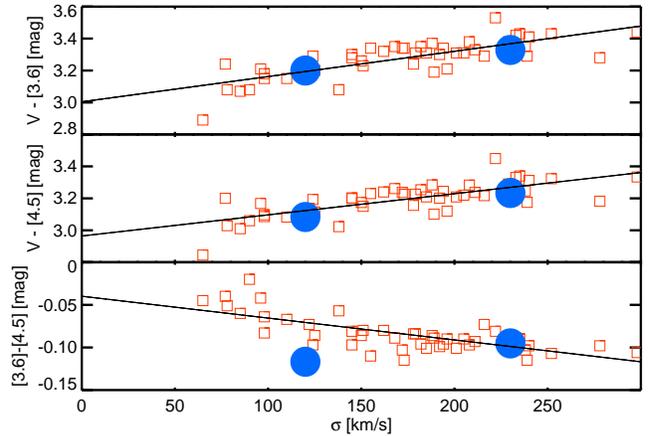}}\qquad
  \caption{Colour versus $\sigma$ visualisations of a selected sample of nearby early-type galaxies from the SAURON survey (red squares). The overplotted large blue dots are the predicted colours from our BaSTI based SSP models for bimodal IMFs of slopes $\mu=1.3$ and $\mu=2.3$, corresponding to the two values of ${\rm \sigma=130 \,km \, s^{-1}}$ and ${\rm \sigma=230 \, km \,s^{-1}}$, respectively. The black straight lines delineate the best fits to the observed galaxies. The upper panel shows the ($V-[3.6]$) colour, the middle one ($V-[4.5]$) and the lower one the \textit{Spitzer} ($[3.6]-[4.5]$) colour.}\qquad
\label{color_sigma_relations}
\end{center}
\end{figure}

\citet{Falcon11b} present tight relations between the ($V-[3.6]$) colour and the central stellar velocity dispersion $\sigma_e$ measured within one effective radius for a sample of elliptical and S0 galaxies from the SAURON survey (see their Figure 7, left bottom panel). In the middle and bottom panels of Figure \ref{color_sigma_relations}, we reproduce the same colour versus $\sigma$ plot for the sample from \citet[][see their Figure 11a]{Peletier12}. Additionally, we overplot the predicted colours from our SSP models for two stellar populations of different $\sigma$. We assume a small $\sigma$ of ${\rm 120 \,km \, s^{-1}}$ and a larger one of ${\rm 230 \,km \, s^{-1}}$, respectively. According to the relation of \citet{Spiniello14}, we choose for the model galaxy of ${\rm \sigma= 120 \, km \, s^{-1}}$ a Kroupa IMF of $\mu \approx 1.3$, whereas a galaxy of ${\rm \sigma = 230 \,km \, s^{-1}}$ is described by a bimodal IMF of $\mu \approx 2.3$. Furthermore, we assumed Solar metallicity and adopted an age of ${\rm6 \, Gyr}$ for the former and of ${\rm 12 \, Gyr}$ for the latter galaxy. Our $V$ band prediction comes from the MILES photometric predictions (see Section \ref{SSP_SED}), since the $V$ filter lies outside the wavelength range covered by our new SSP models. 

From Figure \ref{color_sigma_relations} (upper panel), we can deduce that the location of our model galaxies in the ($V-[3.6])-\sigma$ space is in very good agreement with the data. The same statement holds when plotting the ($V-[4.5]$) colour versus the velocity dispersion $\sigma$, as seen in the middle panel of Figure \ref{color_sigma_relations}. 
However, as seen in the lower panel of Figure \ref{color_sigma_relations}, we do not obtain such a good agreement between the observations and the modelled data points for the \textit{Spitzer} ($[3.6]-[4.5]$) colour versus $\sigma$. 

The reddening of the colours of the galaxies as $\sigma$ decreases is not reproduced well by our SSP models, for which the predicted ($[3.6]-[4.5]$) colour is clearly too blue. This is obviously the same effect as observed before in Section \ref{early-type_galaxies}. There, we explained it by the extended SFHs of the younger, low-mass galaxies which most likely include a significant fraction of thermally pulsating AGB and other young stars. This finding is in accordance with \citet{Falcon11b}, who report a large contribution of these kind of stars to the overall galaxy light in the IR. Contrary to the situation in the optical wavelength range, these types of galaxies apparently cannot be reproduced by a single-burst stellar population model in the IR. Consequently, this wavelength range might enable us for the first time to disentangle the various stellar populations present in young, low-mass, metal-poor galaxies as well as to estimate the contribution of the non-stellar PAH emission. Further investigation on the different stellar populations and non-stellar components in these galaxies will be studied in a future paper. 

\section{Summary and conclusions}

We present new evolutionary single stellar population models covering the so-far poorly studied IR wavelength range beyond the $K$ band, i.e., between ${\rm2.5 \,\mu m}$ and ${\rm 5 \, \mu m}$. Our models use as stellar input 180 carefully selected and prepared spectra from the IRTF stellar library \citep{Cushing05, Rayner09}. We have determined that the resolution of the IRTF spectra is ${\rm 60 \,km \, s^{-1}}$ (see Table \ref{Parameter_coverage}). This makes them very appropriate to study most elliptical galaxies. We determined the atmospheric parameters of the stars as accurately as possible using a compilation from the literature \citep[see][]{Cenarro07} and the colour-temperature relations of \citet{Alonso96, Alonso99}. 

We provide the model spectra in a way which enables the reader to obtain all kinds of ÍR colours by convolving them with the respective filter response functions. Since the stars of the IRTF library are mostly bright, nearby stars of around Solar metallicity, our SSP models are most accurate in the metallicity range between ${\rm[Fe/H]=-0.40}$ and ${\rm[Fe/H]=0.2}$. We also obtained model predictions for slightly higher and lower metallicities, which are, however, of lower quality.\looseness-2


We calculated our models based on both BaSTI \citep{Pietrinferni04} and Padova00 \citep{Girardi00} isochrones. They mainly differ for ages younger than around ${\rm 3 \,Gyr}$, since they use different prescriptions for the AGB star contribution which becomes crucial at these ages. According to \citet{Maraston98, Maraston05}, AGB stars are particularly important at ages between 1 and 2 Gyr. The increasing contribution of this phase results in redder ($[3.6]-[4.5]$) colours.
We cannot extend our models to ages younger than ${\rm1 \,Gyr}$ due to an insufficient number of AGB stars and young, hot stars in the employed IRTF library. A summary of most of the points mentioned so far can be found in Table \ref{Parameter_coverage} in Section \ref{Quality}.

We obtained our SSP models for Kroupa-like, revised Kroupa and unimodal and bimodal IMFs of varying slopes \citep[see][for the different IMF types]{Vazdekis03}. The slopes of the latter two cases range from $\mu=0.3$ to $\mu=3.3$. A unimodal IMF of slope 1.3 resembles a Salpeter-IMF and a bimodal one of the same slope a Kroupa-IMF. As our analysis (see Section \ref{IMF}) shows, our models basically do not depend on the used IMF. Consequently, all the other studies presented in this paper are carried out for models with an underlying Kroupa-like IMF. 

In what follows we summarise the most important results which we obtain from our models. 

\begin{enumerate}

\item We tested the internal consistency of our model predictions by comparing the ($J-K$) colour measured on the IRTF based SSP spectra to the photometric predictions based on extensive photometric libraries published on our webpage (http://miles.iac.es) and found a good agreement, within ${\rm 0.02 \,mag}$ over most of the age range.

\item In general, we can observe only a weak dependence on age and metallicity for our SSP models. The \textit{Spitzer} ($[3.6]-[4.5]$) colour is apart from a very small trend of reddening almost constant with age for ages older than ${\rm \approx 3 \,Gyr}$. In this age range, the small variations in the \textit{Spitzer} ($[3.6]-[4.5]$) colour are almost exclusively driven by metallicity. For younger ages between 1 and ${\rm 3 \,Gyr}$, the ($[3.6]-[4.5]$) colour shows some gradient. It becomes significantly redder depending on the used isochrones (BaSTI or Padova00) due to the enhanced AGB star contribution. Solar and supersolar metallicities result in bluer ($[3.6]-[4.5]$) colours than lower, subsolar ones. This at first sight unexpected behaviour can be explained by the prominent CO absorption band in the ${\rm[4.5] \, \mu m}$ filter which becomes more important for stars of lower temperature and larger metallicity. Hence, for these metallicities, a greater amount of flux is absorbed compared to lower ones causing the observed bluer ($[3.6]-[4.5]$) colour. 
Moreover, the ($[3.6]-[4.5]$) colour is found to be almost unaffected by the underlying IMF.

\item The ($[3.6]-[4.5]$) colours predicted by our SSP models lie in the same range as the observed ones of elliptical galaxies. In particular, for the most massive, older and more metallic, single-burst objects like early-type galaxies, our models provide good fits. However, the younger, lower-mass counterparts show redder colours that might be attributed to a more extended SFH. These galaxies are most likely characterised by the presence of a significant fraction of AGB-dominated intermediate-age stellar populations and of a large amount of PAH emission corresponding to a relatively recent star formation activity. For most of the galaxies showing the reddest ($[3.6]-[4.5]$) colours, enhanced emission in the \textit{Spitzer} ${\rm[8] \, \mu m}$ band which can be attributed to PAHs is indeed observed. Another way to explain these red colours would be the contribution of populations of low metallicities, for which the \textit{Spitzer} colour most likely continues to become redder and redder. In order to confirm or rule out this option, we will need more reliable models in this low-metallicity regime.

\item We obtained $M/L$ ratios in the $K$, ${\rm [3.6] \, \mu m}$ and ${\rm [4.5] \, \mu m}$ bands from the output of our modelling routines. As expected \citep[see, e.g.,][]{Bell01}, the $M/L$ ratio in all these three bands varies much less depending on age and metallicity than when measured in optical filters, since light from the RGB stars dominates in the NIR and in the IR. Due to the small variations in the $(M/L)_{3.6}$ ratio with age and metallicity and thus also with the \textit{Spitzer} ($[3.6]-[4.5]$) colour, \citet{Meidt14} even suggest a parameter-independent $M/L$ ratio of $(M/L)_{3.6}=0.6$. This value is equal to the mean $(M/L)_{3.6}$ from our models when considering the whole age and metallicity range. This result is confirmed by a recent work of \citet{Norris14} who found a variation of the $(M/L)_{3.6}$ ratio within the same limits measured over a comparable age and metallicity range.
Whereas the $(M/L)_{3.6}$ ratio is almost constant as a function of age and metallicity, it depends significantly on the IMF used to calculate the underlying SSP-model. The more bottom-heavy the IMF becomes, the more the $M/L$ ratios increase.

\item Finally, we also compare the behaviour of our SSP models to other models available in the literature for our specific wavelength range. It is important to note that all of them are based on theoretical stellar libraries. 
The \textit{Spitzer} ($[3.6]-[4.5]$) colours derived from the spectra of our SSP models agree almost perfectly with the colour predictions of the models by \citet{Marigo08} and \citet{Bressan12}. Also the agreement with the FSPS models of \citet{Conroy09} is good. The \textit{Spitzer} colours determined from the GALEV \citep{Kotulla09} models show a large offset to the red with respect to the colours measured from all other models mentioned so far. This result can be understood by the use of different theoretical stellar atmospheres and by the non-consideration of the CO absorption affecting the ${\rm[4.5] \, \mu m}$ band.
However, for the Marigo and also for the Bressan models, only colour predictions are available, not SSP spectra. In the case of the other sets of models, spectra can be obtained, but due to their low resolution it is not possible to measure spectral features on them. Thus, our models are the only ones existing up to now which offer the opportunity to obtain line strength indices in the $L$ band. A detailed study of them will be the subject of a following paper. 

\end{enumerate}

We thus present the first models in the wavelength range between 2.5 and ${\rm 5 \, \mu m}$ which are based on an empirical stellar library. Theoretical stellar spectra were only applied in order to fill two gaps in the spectra due to telluric absorption between $\approx \, 2.50$ and $2.80 \,\mu m$ and between $\approx \, 4.20$ and ${\rm 4.50 \,\mu m}$, respectively. The very good agreement of the colour predictions with comparable theoretical models existing in the IR wavelength range between 2.5 and ${\rm 5 \, \mu m}$ as well as the possibility to analyse line strength indices renders our newly developed models utile. Drawbacks are of course the limited range in both age and metallicity due to the limited stellar coverage of the IRTF library and the rather average fit to particularly the reddest and youngest observed data. However, we can state that colours derived from those models by convolution with the respective filter responses can be safely used for all kinds of analyses of observations. 
Our new SSP models will be made available on the MILES webpage (http://miles.iac.es). Future work will encompass a detailed analysis of line strength indices in the NIR and also to a certain extent in the IR as well as combining our models with the existing MIUSCAT models \citep{Vazdekis12} in the optical wavelengths range.

\section*{Acknowledgements}
We thank the anonymous referee who provided us with helpful comments which helped us to improve our paper significantly. B.R. acknowledges financial support to the DAGAL network from the European Union’s Programme FP7/2007-2013/ under REA grant agreement number PITN-GA-2011-289313. JHK acknowledges financial support from the Spanish MINECO under grant number AYA2013-41243-P.

\bibliography{bibliography}

\appendix
\section{Stellar atmospheric parameters}

We present our final set of stellar atmospheric parameters in the following Table \ref{Table_final_SAPs}.

\begin{table*}
\caption{ Final list of stellar atmospheric parameters: The first column indicates the official names of the stars. Columns 2 and 3 contain the derived effective temperatures $T_{\text{eff}}$ and their errors, respectively. In columns 4 and 5 we present the surface gravities $\log(g)$ and their errors. Columns 6 and 7 show the metallicities [Fe/H] and their errors. In column 8 we indicate whether a star contributes with its full weight (100 per cent) or only with 50 per cent or 25 per cent of it. Column 9 reveals how the effective temperatures were compiled. An $\rq  \text{A} \rq$ designates that we made use of the colour-temperature relations of \citet[][see Section \ref{SAPs_color_temperature_relations}]{Alonso96, Alonso99} and a $\rq  \text{CvD} \rq$ of the ones of \citet{Conroy12}, respectively. A $\rq  \text{M} \rq$ means that the respective star was contained in the MILES library, so that we could directly take the value from there. A $\rq  1 \rq$ and a $\rq  + \rq$ express that we obtained the parameters using the approach of \citet[][see also Section \ref{SAPs_literature}]{Cenarro07} based on one or several references, respectively. The abbreviations $\rq  \text{B} \rq$ and $\rq  \text{P} \rq$ mark those stars for which the respective values were taken from the works of \citet{Bergeat01} and \citet{Pickles98}, respectively.}
\label{Table_final_SAPs}
\centering
\begin{minipage}{200mm}
\begin{tabular}{ l c r r c r c c c}
 \hline
  
star name &  $T_{\text{eff}}$ [K]& error [K]& $\log(g)$ [dex]& error [dex]& [Fe/H] [dex] & error [dex] & weight & reference\\
\hline
      HD 31996   & 2290  &	79  &	 $-0.56$  &  0.16   &    0.26  &    0.10  & 100 & B\\
      HD 57160   & 3295  &	195  &	 $-0.46$  &  0.16   &    0.04  &	  0.10  & 100& B \\
      HD 76221   & 2645  &	24 &	 $-0.46$  &  0.16   &   $-0.03$  &	  0.10  & 100& B \\
      HD 44984   & 3035  &	10  &	 $-0.46$  &  0.16   &   $-0.07$  &	0.10 & 100 & B\\
      HD 135153  & 7049  &	47  &	  1.89  &  0.12   &   $-0.79$  &	  0.10 & 100 & 1\\
      HD 13174   & 6422  &	79  &	  3.75  &  0.10   &   $-1.90$  &	  0.05 & 100 & 1\\
      HD 27397   & 7561  &	125 &	  4.05  &  -	  &   $-0.91$  &	  0.10 & 100  & A\\
      HD 173638  & 7155  &	120 &	  1.81  &  0.31   &    0.12  &0.05  & 100& +\\
      BD +382803 & 6597  &	-   &	  1.80  &  -	  &   $-1.92$  &	  0.10 & 100  & P\\
      HD 40535   & 6805  &	-   &	  3.86  &  -	  &   $-0.21$  &	  0.10 & 50 & P\\
     HD 26015   & 6855  &	47  &	  4.35  &  0.12   &    0.17  &	  0.05 & 100 & 1\\
     HD 21770   & 6668  &	125 &	  3.72  &  -	  &   $-0.65$  &	  0.10 & 100 & A\\
     HD 16232   & 6346  &	61  &	  4.54  &  0.18   &    0.03  &	0.09 & 100 & M\\
     HD 87822   & 6533  &	61  &	  4.18  &  0.18   &    0.10  &	  0.09 & 50 & M\\
     HD 213306  & 6051  &	-   &	  1.69  &  -	  &   $-0.12$  &	  0.10  & 100& P\\
     HD 17918   & 6701  &	125 &	  3.80  &  -	  &   $-0.70$  &	  0.10  & 50 & A\\
     HD 27524   & 6622  &	61  &	  4.28  &  0.20   &    0.13  &	  0.10 & 100 & M\\
     HD 11443   & 6305  &	47  &	  3.67  &  0.12   &    0.05  &	  0.10  & 100 & 1\\
     HD 126660  & 6202  &	61  &	  3.84  &  0.18   &   $-0.27$  &	  0.09  & 50&M\\
     HD 51956   & 6200  &	-   &	  1.77  &  -	  &   $-0.09$  &	  0.10 & 100 & P \\
     HD 27383   & 6098  &	61  &	  4.28  &  0.20   &    0.13  &	  0.09  & 100 & M\\
     HD 114710  & 5975  &	85  &	  4.40  &  0.28   &    0.09  &	  0.16 & 100  & M\\
     HD 165908  & 5928  &	85  &	  4.24  &  0.28   &   $-0.53$  &	  0.16  & 100 & M\\
     HD 185018  & 5347  &	120 &	  2.21  &  0.31   &   $-0.24$  &	  0.05  & 100 & +\\
     HD 20619   & 5652  &	61  &	  4.48  &  0.18   &   $-0.28$  &	  0.09  & 100 & M\\
     HD 21018   & 5352  &	125 &	  3.07  &  -	  &    0.00  &	  0.10  & 100 & A\\
     HD 216219  & 5727  &	85  &	  3.36  &  0.28   &   $-0.39$  &	  0.16  & 100& M\\
     HD 10307   & 5838  &	61  &	  4.28  &  0.18   &    0.03  &	  0.09  & 100& M\\
     HD 3421    & 5383  &	120 &	  2.45  &  0.31   &   $-0.17$  &	  0.05 & 100 & +\\
     HD 126868  & 5651  &	120 &	  3.66  &  0.31   &   $-0.02$  &	  0.05  & 100 & +\\
     HD 76151   & 5692  &	61  &	  4.28  &  0.18   &    0.08  &	  0.09 & 100  & M\\
     HD 192713  & 5020  &	125 &	  0.05  &  0.16   &   $-0.41$  &	  0.10 & 100 & A\\
     HD 10697   & 5610  &	120 &	  3.96  &  0.31   &    0.11  &	  0.05 & 100  & +\\
     HD 165185  & 5909  &	28  &	  4.58  &  0.07   &    0.00  &	  0.03 & 100  & 1\\
     HD 202314  & 5007  &	92  &	  1.86  &  0.20   &    0.13  &	  0.06 & 50 & 1 \\
     HD 58367   & 4837  &	54  &	  1.94  &  0.16   &   $-0.14$  &	  0.07  & 100& 1\\
     HD 27277   & 5081  &	125 &	  2.98  &  -	  &   $-0.16$  &	  0.10 & 100 & A\\
     HD 16139   & 5102  &	-   &	  2.90  &  0.16   &   $-0.29$  &	  0.10 & 100  & P\\
     HD 25877   & 5075  &	134 &	  1.88  &  0.18   &    0.09  &	  0.05  & 100 & 1\\
     HD 114946  & 5171  &	61  &	  3.64  &  0.18   &    0.13  &	  0.09 & 100  & M\\
     HD 20618   & 5058  &	121 &	  3.10  &  0.16   &   $-0.26$  &	  0.09 & 100  & 1\\
     HD 135722  & 4847  &	85  &	  2.56  &  0.28   &   $-0.44$  &	  0.16  & 100 &M\\
     HD 75732   & 5079  &	85  &	  4.48  &  0.28   &    0.16  &	  0.16  & 100 & M\\
     HD 170820  & 4604  &	61  &	  1.62  &  0.20   &    0.17  &	  0.10  & 100 & M\\
     HD 9852    & 4648  &	-   &	  2.82  &  -	  &    0.06  &	  0.10 & 100  & P\\
     HD 179870  & 5246  &	125 &	  1.96  &  -	  &    0.11  &	  0.10 & 50  & A\\
     HD 124897  & 4361  &	85  &	  1.93  &  0.28   &   $-0.53$  &	  0.16 & 100  & M\\
     HD 63302   & 4500  &	90  &	  0.20  &  0.20   &    0.12  &	  0.09 & 100  & M\\
     HD 10476   & 5178  &	120 &	  4.43  &  0.31   &   $-0.16$  &	  0.05 & 100  & +\\
     HD 23082   & 4232  &	-   &	  1.59  &  -	  &    0.04  &	  0.10  & 100 &P\\
     HD 212466  & 3797  &	125 &	  0.01  &  -	  &    0.06  &	  0.10 & 50  & A\\
     HD 3765    & 5051  &	120 &	  4.05  &  0.31   &    0.03  &	  0.04 & 100  & +\\
     HD 187238  & 4487  &	56  &	  0.80  &  0.10   &    0.18  &	  0.04 & 100  & 1\\

     \hline
\end{tabular}
\end{minipage}
\end{table*} 

\begin{table*}
\begin{minipage}{200mm}
\contcaption{}
\label{Cenarro}
\centering
\begin{tabular}{ l c r r c r c c c}
 \hline
  
star name &  $T_{\text{eff}}$ [K] & error [K] & $\log(g)$ [dex] & error [dex] & [Fe/H] [dex] & error [dex] & weight & reference\\
\hline 
     HD 16068   & 4296  &	-   &	  1.80  &  0.16   &   $-0.03$  &	  0.10  & 100 & P\\
     HD 207991  & 4152  &	-   &	  2.18  &  -	  &   $-0.29$  &	  0.10 & 100  & CvD\\
     HD 181596  & 3977  &	125 &	  1.89  &  -	  &   $-0.38$  &	  0.10 & 100  & A\\
     HD 36003   & 4465  &	61  &	  4.61  &  0.18   &    0.09  &	  0.10 & 100  & M\\
     HD 194193  & 3785  &	125 &	  1.40  &  0.16   &   $-0.10$  &	  0.10 & 100  & A\\
     HD 209290  & 3650  &	-   &	  4.69  &  -	  &   $-0.15$  &	  0.10 & 100 & CvD\\
     HD 213893  & 3896  &	125 &	  1.62  &  -	  &   $-0.04$  &	  0.10 & 100  & A\\
     HD 19305   & 3972  &	-   &	  4.67  &  -	  &    0.01  &	  0.10 & 100  & CvD\\
     BD +60265  & 3454  &	125 &	  0.38  &  -	  &    0.18  &	  0.10 & 50  & A\\
     HD 36395   & 3737  &	61  &	  4.90  &  0.20   &   $-1.50$  &	  0.09& 100  &M\\
     HD 339034  & 3278  &	125 &	 -0.25  &  -	  &   $-0.07$  &	  0.10 & 50 & A\\
     HD 204724  & 3911  &	125 &	  1.45  &  -	  &   $-0.71$  &	  0.10 & 100  &A\\
     HD 39801   & 3550  &	61  &	  0.05  &  0.16   &    0.03  &	  0.10 & 50 &M\\
     HD 219734  & 3730  &	61  &	  0.90  &  0.18   &    0.27  &	  0.10 & 100 &M\\
     HD 206936  & 3746  &	121 &	 -0.04  &  0.16   &    0.05  &	  0.09 & 50 &1\\
     HD 10465   & 3781  &	125 &	  0.50  &  -	  &   $-0.46$  &	  0.10 & 100 &A\\
     HD 95735   & 3828  &	61  &	  4.90  &  0.20   &   $-0.20$  &	  0.09 & 50 & M\\
     HD 14488   & 3509  &	125 &	 -0.17  &  -	  &   $-0.77$  &	  0.10 & 50 & A\\
     HD 28487   & 3551  &	125 &	  1.03  &  -	  &   $-1.30$  &	  0.10 & 100 & A\\
     HD 14469   & 3551  &	125 &	 -0.16  &  -	  &   $-0.69$  &	  0.10 & 100 & A\\
     Gl 388     & 3304  &	-   &	  4.87  &  -	  &   $-0.16$  &	  0.10 & 100 & P\\
     Gl 268     & 3031  &	-   &	  4.76  &  0.16   &   $-0.16$  &	  0.10 & 50 & P\\
     HD 27598   & 3647  &	125 &	  1.04  &  -	  &   $-0.11$  &	  0.10 & 100 & A\\
     HD 4408    & 3571  &	125 &	  1.00  &  -	  &   $-0.51$  &	  0.10 & 100 & A\\
     Gl 213     & 3030  &	-   &	  5.21  &  -	  &   $-0.54$  &	  0.10 & 100 & CvD\\
     Gl 299     & 3111  &	-   &	  5.23  &  -	  &   $-0.72$  &	  0.10 & 100 & P\\

     HD 18191   & 3250  &	61  &	  0.30  &  0.18   &   $-1.01$  &	  0.10 & 100 & M\\
     HD 196610  & 3060  &	121 &	  0.31  &  0.16   &   $-1.02$  &	  0.09 & 50 & 1\\
     Gl 406     & 2698  &	-   &	  4.63  &  0.16   &   $-0.15$  &	  0.10 & 25 & P\\
     HD 207076  & 2750  &	61  &	 -0.56  &  0.16   &   $-2.61$  &	  0.10 & 100 & M\\

     HD 113139  & 6823  &	120 &	  3.89  &  0.31   &    0.22  &	  0.10 & 100 & +\\
     HD 160365  & 6142  &	47  &	  3.10  &  0.12   &   $-0.23$  &	  0.10 & 100 & 1\\
     HD 108477  & 5172  &	125 &	  3.03  &  -	  &   $-0.01$  &	  0.10 & 100 & A\\
     HD 214850  & 5437  &	-   &	  4.44  &  -	  &   $-0.31$  &	  0.10 & 100 & P\\
     HD 115617  & 5531  &	61  &	  4.32  &  0.18   &   $-0.10$  &	  0.09 & 100 & M\\
     HD 122563  & 4566  &	85  &	  2.81  &  -	  &   $-2.63$  &	  0.16 & 100 & M\\
     HD 101501  & 5401  &	61  &	  4.60  &  0.18   &   $-0.13$  &	  0.09 & 100 & M\\
     HD 145675  & 5264  &	85  &	  4.66  &  0.28   &    0.34  &	  0.16 & 100 & M\\
     HD 2901    & 4459  &	125 &	  2.62  &  -	  &   $-0.03$  &	  0.10 & 100 & A\\
     HD 132935  & 4492  &	-   &	  2.65  &  -	  &   $-0.05$  &	  0.10 & 100 & CvD\\
     HD 221246  & 4228  &	125 &	  2.32  &  -	  &   $-0.24$  &	  0.10 & 100 & A\\
     HD 219134  & 4717  &	85  &	  4.50  &  0.28   &    0.05  &	  0.16 & 100& M\\
    HD 216946  & 3782  &	-   &	  0.74  &  -	  &   $-0.93$  &	  0.10 & 100 & P\\
    HD 237903  & 4070  &	61  &	  4.70  &  0.18   &   $-0.18$  &	  0.10 & 100 & M\\
    HD 42581   & 3731  &	-   &	  4.66  &  -	  &   $-0.16$  &	  0.10 & 50 & CvD\\
    HD 120052  & 3664  &	125 &	  1.26  &  -	  &   $-0.75$  &	  0.10 & 100 & A\\
    HD 40239   & 3349  &	121 &	  0.47  &  0.16   &    0.03  &	  0.09 & 100 & 1\\
    HD 39045   & 3670  &	125 &	  1.12  &  -	  &   $-0.99$  &	  0.10 & 100 & A\\
    Gl 273     & 3108  &	 -  &	  5.00  &  -	  &   $-0.25$  &	  0.10& 100  & CvD\\
    HD 94705   & 3576  &	125 &	  0.77  &  -	  &   $-2.61$  &	  0.10 & 100 & A\\
    HD 108849  & 3434  &	125 &	  0.13  &  0.16   &   $-2.27$  &	  0.10& 100  &A\\
    HD 7927    & 7380  &	92  &	  0.80  &  0.20   &   $-0.25$  &	  0.06 & 100 & 1\\
    HD 6130    & 7041  &	120 &	  2.28  &  0.31   &    0.06  &	  0.05 & 100 & +\\
    HD 89025   & 6953  &	120 &	  3.20  &  0.31   &    0.19  &	  0.09 & 100 & +\\
    HD 108519  & 6952  &	144 &	  4.16  &  -	  &   $-1.51$  &	  0.10 & 100 & A\\
    HD 213135  & 6904  &	144 &	  4.17  &  -	  &   $-0.53$  &	  0.10 & 100 & A\\
    HD 182835  & 7226  &	120 &	  2.12  &  0.31   &    0.09  &	  0.05 & 100 & +\\
    HD 75555   & 6458  &	47  &	  3.63  &  0.12   &   $-0.18$  &	  0.10 & 100 & 1\\
    HD 186155  & 6753  &	47  &	  3.39  &  0.12   &   $-0.25$  &	  0.10 & 100 & 1\\
    HD 218804  & 6261  &	61  &	  4.05  &  0.18   &   $-0.23$  &	  0.09 & 100 & M\\
    HD 215648  & 6167  &	61  &	  4.04  &  0.18   &   $-0.32$  &	  0.09 & 100 & M\\
    HD 201078  & 6157  &	123 &	  2.43  &  -	  &    0.13  &	  0.10 & 100 & M\\
    HD 124850  & 6116  &	61  &	  3.83  &  0.18   &   $-0.10$  &	  0.09 & 100 & M\\

\hline
\end{tabular}
\end{minipage}
\end{table*}    

\begin{table*}
\begin{minipage}{200mm}
\contcaption{}
\label{Cenarro}
\centering
\begin{tabular}{ l c r r c r c c c}

\hline  
star name &  $T_{\text{eff}}$ [K]& error [K] & $\log(g)$ [dex] & error [dex] & [Fe/H] [dex] & error [dex] & weight & reference\\
\hline 
    HD 102870  & 6109  &	85  &	  4.20  &  0.28   &    0.17  &	  0.16 & 100 & M\\
    HD 220657  & 5878  &	47  &	  3.23  &  0.12   &   $-0.65$  &	  0.10 & 100 & 1\\
    HD 111844  & 6122  &	47  &	  3.73  &  0.12   &   $-0.74$  &	  0.10 & 100 & 1\\
    HD 219623  & 6155  &	85  &	  4.17  &  0.28   &   $-0.04$  &	  0.16 & 100 & M\\
    HD 6903    & 5684  &	47  &	  2.75  &  0.12   &   $-0.32$  &	  0.10 & 100 & 1\\
    HD 176051  & 5990  &	47  &	  4.45  &  0.12   &   $-0.18$  &	  0.10 & 100 & 1\\
    HD 109358  & 5868  &	120 &	  4.28  &  0.31   &   $-0.17$  &	  0.05 & 100 & +\\
    HD 74395   & 5250  &	105 &	  1.30  &  0.20   &   $-0.05$  &	  0.09 & 100 & M\\
    HD 95128   & 5813  &	61  &	  4.34  &  0.18   &    0.04  &	  0.09 & 100 & M\\
    HD 39949   & 5240  &	120 &	  1.23  &  0.31   &   $-0.10$  &	  0.05 & 100 & +\\
    HD 42454   & 5238  &	56  &	  1.10  &  0.10   &    0.03  &	  0.04 & 100 & 1\\ 
    HD 219477  & 5635  &	170 &	  2.90  &  0.16   &   $-0.12$  &	  0.10 & 100 & A\\
    HD 176123  & 5221  &	134 &	  2.29  &  0.18   &   $-0.08$  &	  0.05 & 100 & 1\\
    HD 88639   & 5280  &	125 &	  3.08  &  -	  &   $-0.39$  &	  0.10 & 100 & A\\
    HD 94481   & 5302  &	54  &	  2.85  &  0.16   &   $-0.01$  &	  0.08 & 100 & 1\\
    HD 190113  & 4784  &	92  &	  1.67  &  0.20   &    0.05  &	  0.06 & 100 & 1\\
    HD 193896  & 5050  &	170 &	  2.93  &  -	  &   $-0.48$  &	  0.10 & 100 & A\\
    HD 161664  & 4405  &	125 &	  1.63  &  0.16   &   $-0.03$  &	  0.10 & 100 & A\\
    HD 333385  & 4144  &	125 &	  0.08  &  -	  &    0.27  &	  0.10 & 100 & A\\
    HD 182694  & 5057  &	54  &	  2.94  &  0.16   &   $-0.04$  &	  0.08 & 100 & 1\\
    HD 208606  & 4701  &	92  &	  1.19  &  0.20   &    0.12  &	  0.06 & 100 & 1\\
    HD 104979  & 4887  &	120 &	  2.55  &  0.31   &   $-0.43$  &	  0.05 & 100 & +\\
    HD 222093  & 4784  &	125 &	  2.85  &  -	  &    0.08  &	  0.10 & 100 & A\\
    HD 164349  & 4446  &	61  &	  1.50  &  0.18   &    0.39  &	  0.09 & 100 & M\\
    HD 165782  & 4588  &	170 &	  0.54  &  -	  &    0.26  &	  0.10 & 100& A\\
    HD 44391   & 4670  &	120 &	  0.76  &  0.31   &    0.12  &	  0.05 & 100 & +\\
    HD 100006    & 4701  &	120 &	  2.53  &  0.31   &   $-0.17$  & 0.05 & 100 & +\\
    HD 91810   & 4604  &	125 &	  2.74  &  -	  &   $-0.01$  &	  0.10 & 100 & A\\
    HD 36134   & 4677  &	125 &	  2.77  &  -	  &   $-0.29$  &	  0.10 & 100 & A\\
    HD 25975   & 4951  &	121 &	  3.44  &  0.16   &    0.05  &	  0.09 & 100 & 1\\
    HD 142091  & 4788  &	120 &	  3.23  &  0.31   &   $-0.00$  &	  0.05 & 100 & +\\
    HD 165438  & 4862  &	97  &	  3.40  &  0.20   &    0.02  &	  0.10 & 100 & M\\
    HD 137759  & 4498  &	61  &	  2.38  &  0.18   &    0.05  &	  0.09 & 100 & M\\
    HD 114960  & 4301  &	125 &	  2.29  &  -	  &   $-0.09$  &	  0.10 & 100&  A\\
    HD 99998   & 3939  &	61  &	  1.80  &  0.18   &   $-0.31$  &	  0.09 & 100 & M\\
    HD 35620   & 4367  &	85  &	  1.75  &  0.28   &   $-0.03$  &	  0.16 & 100 & M\\
    HD 178208  & 4473  &	170 &	  2.35  &  -	  &   $-0.23$  &	  0.10 & 100 & A\\
    HD 185622  & 3848  &	-   &	  0.28  &  0.16   &   $-0.15$  &	  0.10 & 100 & P\\
    HD 201065  & 3879  &	125 &	  1.14  &  0.16   &   $-0.49$  &	  0.10 & 100 & A\\
    HD 45977   & 4657  &	34  &	  4.27  &  0.07   &    0.01  &	  0.05 & 100 & 1\\
    HD 120477  & 4092  &	55  &	  1.50  &  0.16   &   $-0.56$  &	  0.08 & 100 & 1\\
    HD 3346    & 3909  &	170 &	  1.19  &  0.16   &   $-0.00$  &	  0.10 & 100 & A\\
    HD 181475  & 4024  &	79  &	  1.40  &  0.11   &    0.00  &	  0.06 & 100 &1\\
    HD 201092  & 3843  &	120 &	  4.50  &  0.11   &   $-0.41$  &	  0.06 & 100 &+\\
    HD 164136  & 6799  &	340 &	  3.96  &  -	  &   $-0.30$  &	  0.10& 50 &  M\\
    HD 236697  & 3594  &	125 &	  0.52  &  -	  &    0.18  &	  0.10 & 50 & A\\
    HD 35601   & 3617  &	120 &	  0.06  &  0.31   &   $-0.20$  &	  0.05& 100  &+\\
    HD 14404   & 3640  &	125 &	  0.32  &  -	  &   $-0.07$  &	  0.10& 100  & A\\
    HD 23475   & 3863  &	170 &	  0.78  &  -	  &   $-0.63$  &	  0.10& 100  & A\\
    Gl 806     & 3611  &	-   &	  4.71  &  -	  &   $-0.19$ &	  0.10& 100  & CvD\\
    HD 204585  & 3428  &	 -  &	  1.00  &  -	  &   $-1.46$  &	  0.10& 100  & CvD\\
    HD 19058   & 3544  &	125 &	  0.99  &  -	  &   $-0.96$  &	  0.10& 100  & A\\
    HD 214665  & 3397  &	125 &	  0.91  &  -	  &   $-1.40$  &	  0.10& 100  & A\\
    HD 156014  & 3236  &	-   &	  0.73  &  0.16   &   $-2.61$  &	  0.10& 100  & P\\
    HD 175865  & 3420  &	61  &	  0.50  &  0.18   &    0.14  &	  0.10 & 100 & M\\
    Gl 51	   & 2794  &	-   &	  5.36  &  -	  &   $-0.15$  &	  0.10& 100  & CvD\\
    Gl 866     & 2951  &	-   &	  5.27  &  -	  &   $-0.15$  &	  0.10& 100  & P\\
   
    HD 190323  & 5715  &	56  &	  0.40  &  0.10   &   $-0.24$  &	  0.04 & 100 & 1\\
    HD 18474   & 4899  &	120 &	  2.36  &  0.31   &   $-0.30$  &	  0.05& 100  & +\\
    HD 92055   & 2965  &	57  &	 -0.56  &  0.16   &   $-0.06$  &	  0.10& 100  & B\\
    HD 76846   & 4687  &	125 &	  2.10  &  0.16   &   $-0.11$  &	  0.10& 100  &A\\
       Gl 381  & 3319  &	-   &	  4.64  &  -	  &    $-0.00$     &	  - & 100 & CvD\\
       Gl 581  & 3197  &	-   &	  4.57  &  -	  &    $-0.00$     &	  -  & 100 & CvD\\
    HD 179821 & 4706  &    -  &	0.34  &  -   &       $-0.00$	 &    - & 100 & A\\ 
\hline
\end{tabular}
\end{minipage}
\end{table*}

\end{document}